\documentclass[11pt]{article}
\usepackage{amsfonts}
\usepackage{amssymb}
\usepackage{amsmath}
\usepackage{amsxtra}
\usepackage{graphicx}
\usepackage{psfrag}
\usepackage{mathrsfs}
\usepackage{stmaryrd}
\usepackage{subfigure}
\usepackage{tikz}
\usepackage{empheq}
\usepackage[normalem]{ulem}
\usepackage{pdfpages}
\usepackage{color}
\usepackage[superscript]{cite}
\usepackage{longtable}
\usepackage{footmisc}

\begin{document}
\def\BGamma{\mbox{\boldmath$\Gamma$}}
\def\BDelta{\mbox{\boldmath$\Delta$}}
\def\BTheta{\mbox{\boldmath$\Theta$}}
\def\BLambda{\mbox{\boldmath$\Lambda$}}
\def\BXi{\mbox{\boldmath$\Xi$}}
\def\BPi{\mbox{\boldmath$\Pi$}}
\def\BSigma{\mbox{\boldmath$\Sigma$}}
\def\BUpsilon{\mbox{\boldmath$\Upsilon$}}
\def\BPhi{\mbox{\boldmath$\Phi$}}
\def\BPsi{\mbox{\boldmath$\Psi$}}
\def\BOmega{\mbox{\boldmath$\Omega$}}
\def\Balpha{\mbox{\boldmath$\alpha$}}
\def\Bbeta{\mbox{\boldmath$\beta$}}
\def\Bgamma{\mbox{\boldmath$\gamma$}}
\def\Bdelta{\mbox{\boldmath$\delta$}}
\def\Bepsilon{\mbox{\boldmath$\epsilon$}}
\def\Bzeta{\mbox{\boldmath$\zeta$}}
\def\Beta{\mbox{\boldmath$\eta$}}
\def\Btheta{\mbox{\boldmath$\theta$}}
\def\Biota{\mbox{\boldmath$\iota$}}
\def\Bkappa{\mbox{\boldmath$\kappa$}}
\def\Blambda{\mbox{\boldmath$\lambda$}}
\def\Bmu{\mbox{\boldmath$\mu$}}
\def\Bnu{\mbox{\boldmath$\nu$}}
\def\Bxi{\mbox{\boldmath$\xi$}}
\def\Bpi{\mbox{\boldmath$\pi$}}
\def\Brho{\mbox{\boldmath$\rho$}}
\def\Bsigma{\mbox{\boldmath$\sigma$}}
\def\Btau{\mbox{\boldmath$\tau$}}
\def\Bupsilon{\mbox{\boldmath$\upsilon$}}
\def\Bphi{\mbox{\boldmath$\phi$}}
\def\Bchi{\mbox{\boldmath$\chi$}}
\def\Bpsi{\mbox{\boldmath$\psi$}}
\def\Bomega{\mbox{\boldmath$\omega$}}
\def\Bvarepsilon{\mbox{\boldmath$\varepsilon$}}
\def\Bvartheta{\mbox{\boldmath$\vartheta$}}
\def\Bvarpi{\mbox{\boldmath$\varpi$}}
\def\Bvarrho{\mbox{\boldmath$\varrho$}}
\def\Bvarsigma{\mbox{\boldmath$\varsigma$}}
\def\Bvarphi{\mbox{\boldmath$\varphi$}}
\def\bone{\mbox{\boldmath$1$}}
\def\bzero{\mbox{\boldmath$0$}}
\def\bnabla{\mbox{\boldmath$\nabla$}}
\def\bvarepsilon{\mbox{\boldmath$\varepsilon$}}
\def\bA{\mbox{\boldmath$ A$}}
\def\bB{\mbox{\boldmath$ B$}}
\def\bC{\mbox{\boldmath$ C$}}
\def\bD{\mbox{\boldmath$ D$}}
\def\bE{\mbox{\boldmath$ E$}}
\def\bF{\mbox{\boldmath$ F$}}
\def\bG{\mbox{\boldmath$ G$}}
\def\bH{\mbox{\boldmath$ H$}}
\def\bI{\mbox{\boldmath$ I$}}
\def\bJ{\mbox{\boldmath$ J$}}
\def\bK{\mbox{\boldmath$ K$}}
\def\bL{\mbox{\boldmath$ L$}}
\def\bM{\mbox{\boldmath$ M$}}
\def\bN{\mbox{\boldmath$ N$}}
\def\bO{\mbox{\boldmath$ O$}}
\def\bP{\mbox{\boldmath$ P$}}
\def\bQ{\mbox{\boldmath$ Q$}}
\def\bR{\mbox{\boldmath$ R$}}
\def\bS{\mbox{\boldmath$ S$}}
\def\bT{\mbox{\boldmath$ T$}}
\def\bU{\mbox{\boldmath$ U$}}
\def\bV{\mbox{\boldmath$ V$}}
\def\bW{\mbox{\boldmath$ W$}}
\def\bX{\mbox{\boldmath$ X$}}
\def\bY{\mbox{\boldmath$ Y$}}
\def\bZ{\mbox{\boldmath$ Z$}}
\def\ba{\mbox{\boldmath$ a$}}
\def\bb{\mbox{\boldmath$ b$}}
\def\bc{\mbox{\boldmath$ c$}}
\def\bd{\mbox{\boldmath$ d$}}
\def\be{\mbox{\boldmath$ e$}}
\def\bff{\mbox{\boldmath$ f$}}
\def\bg{\mbox{\boldmath$ g$}}
\def\bh{\mbox{\boldmath$ h$}}
\def\bi{\mbox{\boldmath$ i$}}
\def\bj{\mbox{\boldmath$ j$}}
\def\bk{\mbox{\boldmath$ k$}}
\def\bl{\mbox{\boldmath$ l$}}
\def\bm{\mbox{\boldmath$ m$}}
\def\bn{\mbox{\boldmath$ n$}}
\def\bo{\mbox{\boldmath$ o$}}
\def\bp{\mbox{\boldmath$ p$}}
\def\bq{\mbox{\boldmath$ q$}}
\def\br{\mbox{\boldmath$ r$}}
\def\bs{\mbox{\boldmath$ s$}}
\def\bt{\mbox{\boldmath$ t$}}
\def\bu{\mbox{\boldmath$ u$}}
\def\bv{\mbox{\boldmath$ v$}}
\def\bw{\mbox{\boldmath$ w$}}
\def\bx{\mbox{\boldmath$ x$}}
\def\by{\mbox{\boldmath$ y$}}
\def\bz{\mbox{\boldmath$ z$}}
\newcommand*\mycirc[1]{%
  \begin{tikzpicture}
    \node[draw,circle,inner sep=1pt] {#1};
  \end{tikzpicture}
}
\newcommand{\upcite}[1]{\textsuperscript{\textsuperscript{\cite{#1}}}}

\makeatletter
\def\@biblabel#1{#1.}
\makeatother


\title{Intercalation driven porosity effects in coupled continuum models for the electrical, chemical, thermal and mechanical response of battery electrode materials}
\author{Z. Wang\thanks{Department of Mechanical Engineering, University of Michigan}, J. Siegel\thanks{Department of Mechanical Engineering, University of Michigan}, \& K. Garikipati\thanks{Departments of Mechanical Engineering, and Mathematics, University of Michigan, corresponding author, {\tt krishna@umich.edu}}}
\maketitle

\begin{abstract}
We present a coupled continuum formulation for the electrostatic, chemical, thermal and mechanical processes in battery materials. Our treatment applies on the macroscopic scale, at which electrodes can be modelled as porous materials made up of active particles held together by binders and perfused by the electrolyte. Starting with the description common to the field, in terms of reaction-transport partial differential equations for ions, variants of the classical Poisson equation for electrostatics, and the heat equation, we add mechanics to the problem. Our main contribution is to model the evolution of porosity as a consequence of strains induced by intercalation, thermal expansion and mechanical stresses. Recognizing the potential for large local deformations, we have settled on the finite strain framework. \textcolor{black}{We present a detailed computational study of the influence of the dynamically evolving porosity, upon ion distribution, electrostatic potential fields, charge-discharge cycles and mechanical force generated in the cell.}
\end{abstract}
%
%
\section{Introduction}
The intercalation of lithium, thermal and mechanical strains drive volume changes in the active material of battery electrodes. The lattice-scale distortions induced by intercalation change the kinetics of lithium transport. At a larger scale, as the particles deform, the porous microstructure of the composite electrode also evolves, and can have a pronounced effect on the effective conductivity, diffusivity and reaction rates throughout the cell. On a solely theoretical basis, the physics suggests that there will be changes in the electrochemical response of the cell as a consequence of mechanics.

The literature on battery materials has seen a number of recent works that explore some of these effects. \textcolor{black}{Cannarella et al.\cite{ArnoldCannarella2014JES} showed that separators stiffen mechanically under intercalation-induced compressive stresses in the electrodes of cells that were also loaded externally.} This stiffening is a consequence of the nonlinearly elastic response of polymer separators. Gor et al.\cite{ArnoldGor2014JES} developed a model for the variation of mechanical properties due to such compression. Shi et al.\cite{Xiao2011JPS} and Xiao et al.\cite{Xiao2010JPS} used a linear relationship between the local strain of swelling due to intercalation and the stress. Mendoza et al.\cite{Roberts2016} used a similar linear relationship for the lithiation-induced swelling. \textcolor{black}{The local stress induced in the compressible separator is known to cause nonuniform lithium transport\cite{ArnoldCannarella2014FadeJPS}. The loss of active material in the electrodes due to particle fracture, growth of the interface layer, and disconnection of the electronic pathways leads to capacity fade with cycling \cite{White2008JPS}. }

Studies\cite{Wood2014,Klingele2015,Thiele2014,Thiele2015} on the changing porosity and microstructure of electrodes and their effect on transport have been carried out by reconstructions of the full 3-D geometry based on X-ray tomography data. Numerical investigations also have been carried out for the effect of porosity on transport parameters. \textcolor{black}{These include the work of Kehrwald et al.\cite{Kehrwald2011JES} who used tomography-based reconstructions of the  microstructure of porous electrodes to investigate the influence of tortuosity. Du et al.\cite{Du2014JES} conducted microscale simulations on microstructures modelled as random packings of ellipsoidal particles to determine the effective diffusivity and conductivity used in macroscopic, homogenized electrode scale models.} Also related are the mass transport simulations\cite{Wieser2015JPS}  on solid/void-resolving voxel meshes, derived from image-processing of active particle microstructures, to determine the effective diffusivity. Earlier,  Stephenson et al.\cite{Wheeler2007JES} had developed an ``inter-particle'' model to account for different particle sizes and material conductivities in porous electrodes. However, these authors ignored the varying microstructural geometry and the resulting transient volume changes in their simulations.

The literature has many reports of particle expansion and porosity variation  due to lithiation during battery operation\cite{Lambros2014JPS,Nagpure2013JPS,Ebner2013,Brandon2010EC}. Channagiri et al.\cite{Nagpure2013JPS} and Ebner et al.\cite{Ebner2013} also show that, as a result, the porosity can be non-uniform through the thickness of the electrode. However most battery performance simulations assume that the particle size of the active material, porosity and thickness of the electrode remain constant. \textcolor{black}{A few studies have been carried out with parametric variation of these quantities across simulations\cite{White2000JPS,White2010JES}. However, the coupled physics that drives the variation of these quantities during battery operation has remained beyond the scope of these studies.} Rieger et al.\cite{Rieger2016JES} modelled the expansion of active material particles due to intercalation by assuming the stress to be linearly dependent on lithium concentration. The authors extended this model to the electrode by maintaining a constant volume fraction, i.e., constant porosity of the particles. However, if the particles swell with intercalation, their volume fraction increases, i.e., the porosity decreases, as has been reported \textcolor{black}{elsewhere\cite{Lambros2014JPS,Nagpure2013JPS,Ebner2013,Brandon2010EC}, but not modelled by Rieger et. al.\cite{Rieger2016JES}}  Awarke et al.\cite{Awarke2011} studied the variation of porosity in microscale computations on representative volume elements by assuming a porosity expression involving the local state of charge and volumetric strain. Then, with a homogenized model at  the macroscopic, electrode scale, they imposed spatial profiles for the state of charge and studied how the porosity and conductivity varied. 

To the best of our knowledge, however, there have not been modelling studies \textcolor{black}{combining the following features: (a) Lithium concentration fields that evolve in space and time under the governing partial differential equations for electro-chemical charge transport, (b)  temperature fields that evolve under the full governing partial differential equation for heat production and transport, (c) the lithium concentration and temperature fields that drive strain fields governed by the {\color{black}partial differential equations of mechanics} for nonlinear elasticity, and cause space- and time-varying porosity changes, which (d) close the loop by inducing variations in conductivity, diffusivity and reaction rates.} In this communication we aim to fill this gap in the models. We first present  such a framework that constitutes an extension of the pseudo-two-dimensional model of Doyle et al.\cite{Moyle1993}, and accounts for non-constant porosity and particle size in the setting of finite strain elasticity (Section \ref{sec:EleMeModel}). We briefly discuss numerical and computational issues (Section \ref{sec:numerics}), before proceeding to a  study of porosity effects on battery performance with and without thermal effects (Section \ref{sec:numericalexamples}). Concluding remarks appear in Section \ref{sec:conclusions}.

\section{The coupled electro-chemo-thermo-mechanical model}
\label{sec:EleMeModel}
%
\begin{figure}[hbtp]
\centering
\includegraphics[scale=0.35]{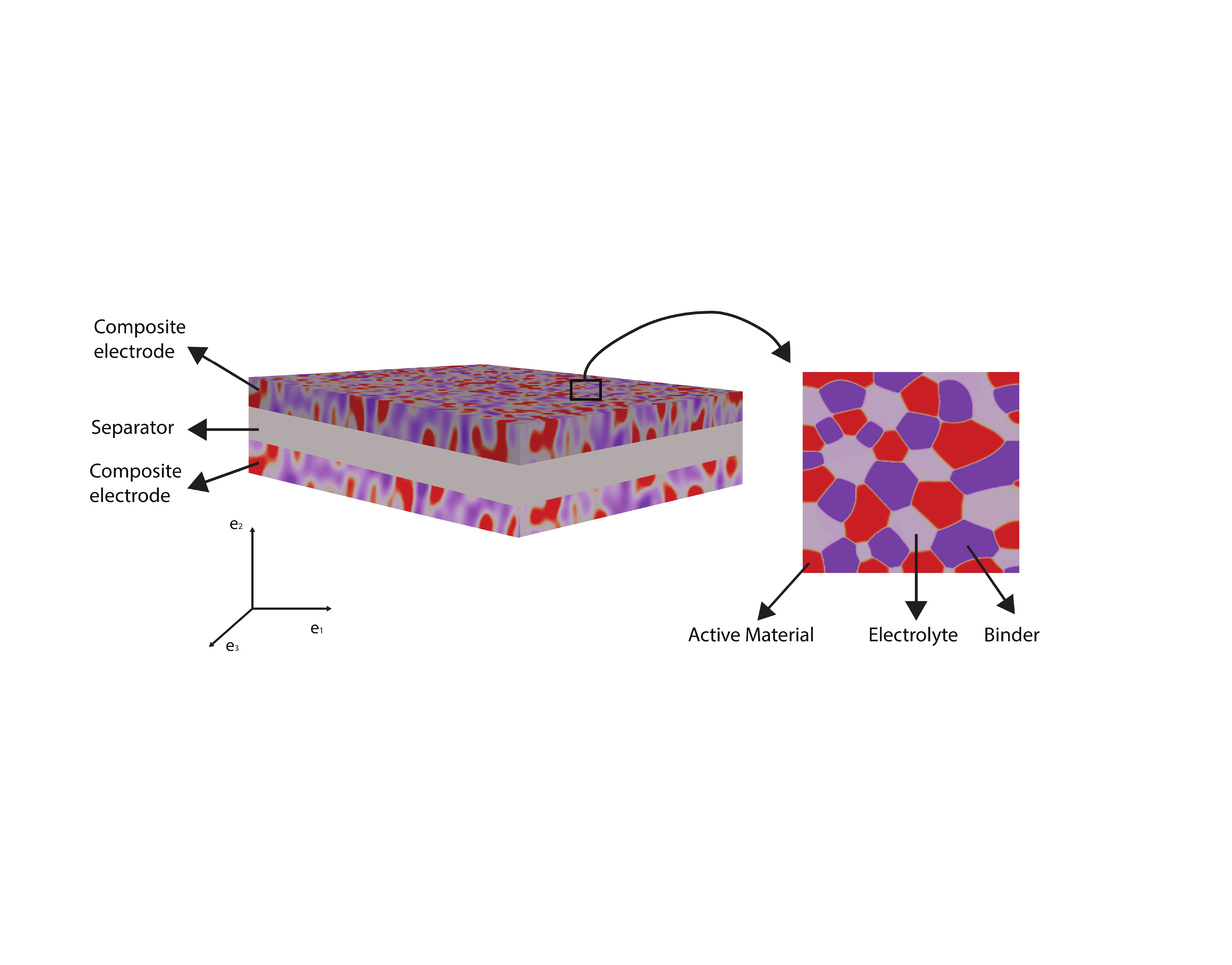}
\caption{A schematic of the three-dimensional cell showing porous electrodes and separator. In this model, the electrolyte fills the pores. The current collectors are not shown.}
\label{fig:3DporousBattery}
\end{figure}
We lay down the governing electrochemical equations, while accounting for changes in configuration induced by the finite strain kinematics in three dimensions. This treatment may be viewed as an extension of the pioneering work of  Newman and Tiedemann\cite{Newman1975}. \textcolor{black}{It is then coupled with the thermal field governed by the heat equation, with heat production from charge transport and reactions.} The novel aspect of this framework is the incorporation of mechanics at finite strain driven by lithiation- and temperature-induced swelling, and the {\color{black}spatio-temporal} evolution of porosity, particle size and maximum allowable concentrations based on the corresponding kinematics. The effect of porosity on transport and electrostatic coefficients completes the coupled formulation.

\subsection{Electro-chemo-thermal equations}
A battery cell usually consists of porous, positive and negative electrodes, a separator and a current collector (see Figure \ref{fig:3DporousBattery}). The simplified porous electrode model\cite{Moyle1993} is widely used,  based on the theory set forth by Newman and Tiedemann\cite{Newman1975}. {\color{black}The equations that follow correspond to those in Newman and Tiedemann\cite{Newman1975}. They are modified to account first for the deformed configuration of the electrode. Since the evolution of porosity depends on the kinematics of deformation, these effects are treated jointly in Section \ref{sec:porosity}.} We note that the porosity was assumed to be constant by Doyle et al.\cite{Moyle1993}. All equations are expressed in the deformed configuration. We only present the final equations here. Detailed derivations appear in the Appendix.
 
\subsubsection{The electrochemical equations for finitely deforming electrodes}
In the deformed configuration of the electrode, $\Omega_\text{e}$, the integral form of mass balance for lithium is
\begin{align}
\int_{\Omega_\text{e}}\frac{\partial (\epsilon_\text{s}C_{1})}{\partial t}\mathrm{d}v+\int_{\Omega_\text{e}}\epsilon_\text{s}{a_\text{p}}j_n\mathrm{d}v=0,
\end{align}
where $C_1$ is the concentration of lithium in the deformed configuration, $\epsilon_\text{s}$ is the volume fraction of solid particles, and $a_\text{p}$ is related to the inverse radius. For spherical particles of radius $R_\text{p}$, it is defined as $a_\text{p} = 3/R_\text{p}$. Finally, $j_n$ is the normal flux of lithium on the particle's surface. A standard localization argument leads to the ordinary differential equation for mass balance of lithium:
\begin{align}
\frac{\partial}{\partial t}({\epsilon_\text{s}C_{1}})+\epsilon_\text{s}{a_\text{p}}j_n=0,\quad \mathrm{in} \;\Omega_\text{e}.
\label{eq:conserC1Strong}
\end{align}

The integral form of mass balance  for $\text{Li}^+$ ions over the deformed configuration of the electrode is
\begin{align}
\int_{\Omega_\text{e}}\frac{\partial (\epsilon_\text{l}C_{2})}{\partial t}\mathrm{d}v=\int_{\Omega_\text{e}}\nabla \cdot(\epsilon_\text{l}D_\text{eff}\nabla C_{2}) \mathrm{d}v+(1-t^0_+)\int_{\Omega_\text{e}}\epsilon_\text{s}{a_\text{p}}j_n \mathrm{d}v
\end{align}
where $C_2$ is the concentration of $\text{Li}^+$ ions in the deformed configuration, $\epsilon_\text{l}$ is the volume fraction of pores in the electrode, $t^0_+$ is the
lithium ion transference number, and $D_\text{eff}$ is the effective diffusivity {\color{black}which depends on the porosity approximated by the Bruggeman relationship as shown in Equation (\ref{pa:Deff})}. A standard localization argument leads to the partial differential equation for mass balance of $\text{Li}^+$ ions:
\begin{align}
\frac{\partial }{\partial t}(\epsilon_\text{l}C_{2})=\nabla\cdot (\epsilon_\text{l}D_\text{eff}\nabla C_{2})+(1-t^0_+)\epsilon_\text{s}{a_\text{p}}j_n.
\label{eq:conserC2Strong}
\end{align}
In widely used porous electrode models\cite{Moyle1993}, the porosity (volume fraction) is assumed to be constant during battery operation, and mechanical deformation is neglected. Here, we also assume low material velocities and volumetric rates of deformation, although we proceed to model the change in porosity due to the deformation; i.e., we only assume rates to be small, while the deformation itself is finite.  These steps and the corresponding arguments appear in the Appendix. Also note that the concentrations $C_1$ and $C_2$ are properly defined with respect to the deformed configuration; i.e., per unit deformed volume.

The equations for the electric fields following Pollard and Newman\cite{Pollard1980} and Doyle et al.\cite{Moyle1993} are,
{\color{black}
\begin{equation}
\nabla\cdot\left(\gamma_\text{eff}(-\nabla\phi_\text{E})+\gamma_\text{eff}\frac{2R\theta}{F}(1-t^{0}_{+})\nabla\ln C_{2}\right)=a_\text{p}Fj_{n}
\label{eq:phieEq}
\end{equation}
}
\begin{equation}
\nabla\cdot\left(\sigma_\text{eff}(-\nabla\phi_\text{S})\right)=-a_\text{p}Fj_{n}
\label{eq:phisEq}
\end{equation}
where $\phi_\text{E}$ and $\phi_\text{S}$ are, respectively, the electric potential fields in the electrolyte and solid, {\color{black}$\gamma_\text{eff}$} and $\sigma_\text{eff}$ are the corresponding {\color{black}effective} conductivities {\color{black}which depend on the porosity as shown in Equation (\ref{pa:Keff}) and (\ref{pa:sigmaeff})}, $R$ is the universal gas constant and $\theta$ is the temperature. 

\begin{figure}[hbtp]
\centering
\includegraphics[scale=0.35]{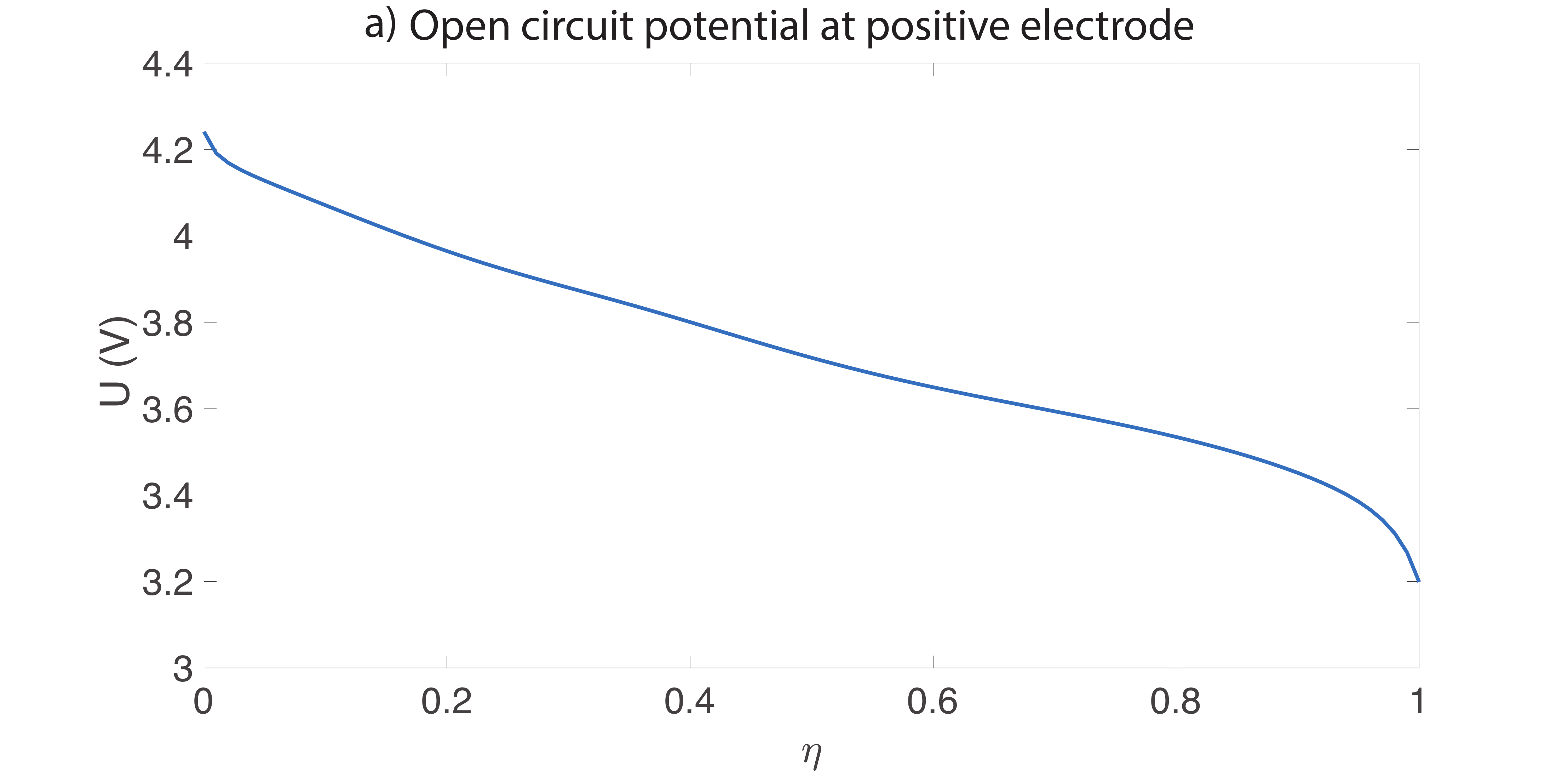}
\includegraphics[scale=0.35]{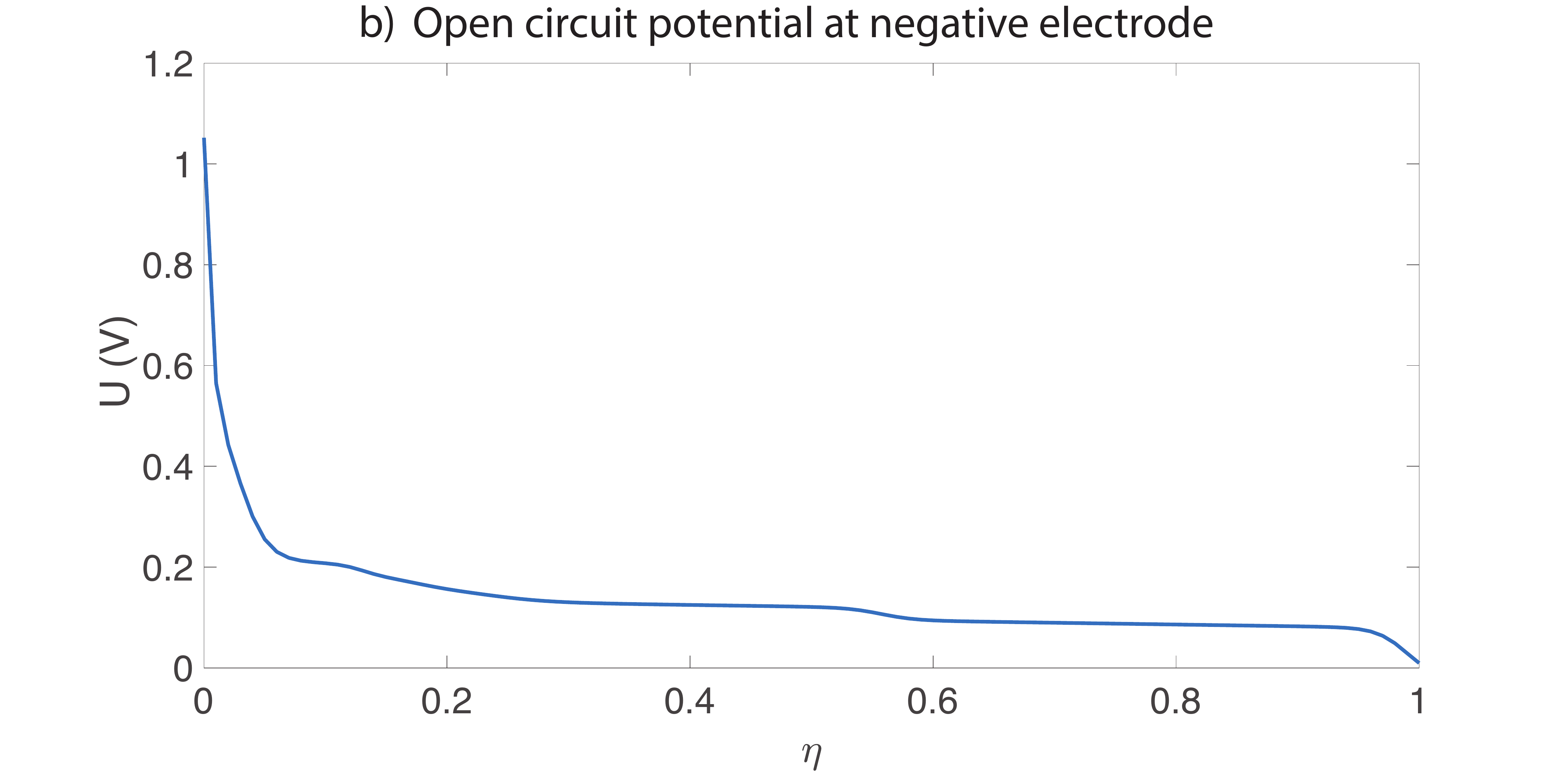}
\caption{Open circuit potential $U$, as a function of $\eta = C_1/C_1^\text{max}$}
\label{fig:U}
\end{figure}

The electrochemical equations are completed with specification of the Butler-Volmer model for the flux of lithium, $j_n$:
\begin{align}
j_n&=j_0\left(\text{exp}\left(\frac{\alpha_aF}{R\theta}(\phi_\text{S}-\phi_\text{E}-U)\right)-\text{exp}\left(-\frac{\alpha_aF}{R\theta}(\phi_\text{S}-\phi_\text{E}-U)\right)\right) \label{eq:BVeuqation}\\
j_0&=k_0(C_2)^{\alpha_a} \frac{(C_1^{\text{max}}-C_{1_\text{surf}})^{\alpha_a}}{(C_1^{\text{max}})^{\alpha_a}}  \frac{(C_{1_\text{surf}})^{\alpha_c}}{{(C_1^\text{max}})^{\alpha_c}}
\label{eq:BVeuqationj0}
\end{align}
where $\alpha_a, \alpha_c$ are transfer coefficients, {\color{black}$k_0$ is a kinetic rate constant, $C_{1_\text{surf}}$ is the value of $C_1$ at the particle surface given by equation (\ref{eq:C1surface}) and $C^{\text{max}}_1$ is the maximum concentration of lithium that the particle can contain. Since the particle volume varies due to deformation, $C^{\text{max}}_1$ is not constant but is defined by Equation (\ref{eq:Cmax}) as discussed in Section \ref{sec:revisedCmax}. {\color{black}The number of moles of electrode host material remains constant in the whole electrode, but as the electrode swells the concentration decreases.} Upon defining $\eta=\frac{C_1}{C^{\text{max}}_1}$, the open circuit potential  $U(\eta)$ can be written as a fit with the following forms, and is shown in Figure \ref{fig:U}:
\begin{align}
U(\eta)=\left.
\begin{cases}
\frac{-0.0923-7.87\eta+50.07\eta^2-122.28\eta^3+82.98\eta^4+140.29\eta^5-374.73\eta^6+403.25\eta^7-221.19\eta^8+49.33\eta^9}{-0.02-1.9\eta+11.73\eta^2-28.78\eta^3+27.54\eta^4-8.63\eta^5} &\text{+ve electrode}\\
 & \\
0.266+0.555e^{-178.97\eta}-0.012\tanh(\frac{\eta-0.557}{0.028})-0.0117\tanh(\frac{\eta-0.239}{0.049})&\\
-0.0129\tanh(\frac{\eta-0.175}{0.035})-0.05\tanh(\frac{\eta-0.99}{0.0245})-0.035\eta-0.012\tanh(\frac{\eta-0.13}{0.02})&\\
-0.152\tanh(\frac{\eta-0.03}{0.023}) &\text{-ve electrode}
\end{cases}
\right.
\label{eq:BVeuqationU}
\end{align}
}

{\color{black}The half cell voltages were fit based on experimental data collected using electrodes harvested from a commercial NMC-graphite cell and cycled at the C/100 hour rate vs lithium metal.}

\subsubsection{The standard thermal equations}
\textcolor{black}{Heat generation and transport are governed by the heat equation, which is derived from the first law of thermodynamics.} For the temperature $\theta$, we have the standard form of the heat equation in the electrodes\cite{Bernardi1985}:
\begin{align}
\rho C_p\frac{\text{d}\theta}{\text{d}t}=\lambda \nabla^2\theta+Q_{\text{rxn}}+Q_{\text{rev}}+Q_{\text{ohm}}
\label{eq:ThermalElectrode}
\end{align}
where $\rho$ is  the mass density of the electrode,  $C_p$ is specific heat and $\lambda$ is the thermal conductivity.  In the separator, we have:
\begin{align}
\rho C_p\frac{\text{d}\theta}{\text{d}t}=\lambda \nabla^2\theta+Q_{\text{ohm}}.
\label{eq:ThermalSeparator}
\end{align}
The heat generation terms are:
\begin{subequations}
\begin{alignat}{2}
Q_{\text{rxn}}&=Faj_n(\phi_\text{S}-\phi_\text{E}-U)  &&\;\text{irreversible entropic heat} \label{pa:irreversibleentropicheat}\\
Q_{\text{rev}}&=Faj_n\theta\frac{\partial U}{\partial \theta} &&\;\text{reversible entropic heat} \label{pa:reversibleelectrochemicalreactionheat}\\
Q_{\text{ohm}}&=-\bi_1\cdot\nabla\phi_\text{S}-\bi_2\cdot\nabla\phi_\text{E} &&\;\text{{\color{black}Joule heating in electrode} } \label{JouleheatElectrode)}\\
Q_{\text{ohm}}&=-\bi_2\cdot\nabla\phi_\text{E} &&\;\text{{\color{black}Joule heating in separator}} \label{JouleheatSeparator)}
\end{alignat}
\end{subequations}
{\color{black}The function $\partial U/\partial \theta$ in Equation (\ref{pa:reversibleelectrochemicalreactionheat}) was also written in terms of $\eta$ and fit to data\cite{KiYong2014}. The fit appears in the Appendix.}
Of the full set of electro-chemo-thermal equations (\ref{eq:conserC1Strong}), (\ref{eq:conserC2Strong}) and (\ref{eq:phieEq}-\ref{eq:ThermalElectrode}) hold in the electrode, and equations (\ref{eq:conserC2Strong}), (\ref{eq:phieEq}) and (\ref{eq:ThermalSeparator}) hold in the separator.

Concentration-, temperature- and porosity-dependent constitutive functions are listed below. All other coefficients are summarized in the Appendix. {\color{black}The conductivity\cite{Kim2011} and diffusivity\cite{Renganathan2010} of the electrolyte appear in Equations (\ref{pa:Ke}) and (\ref{pa:Dl}).  The Bruggeman relation\cite{Moyle1993}, which appears in Equations (\ref{pa:Keff}--\ref{pa:sigmaeff}), is used to calculate the effective conductivity and diffusivity in the porous electrode.} 
{\color{black}
\begin{align}
\gamma=((34.5\exp(-798/\theta)(1.0\times10^3 C_2)^3-485\exp(-1080/\theta)(1.0\times10^3C_2)^2 \nonumber \\
+2440\exp(-1440/\theta))(1.0\times10^3 C_2))/10)\times10^6  \quad \sim p\Omega^{-1}/\mu\text{m}
\label{pa:Ke}
\end{align}
}
\begin{align}
D=\left(10^{-4.43-\frac{54}{\theta-5.0\times10^3 C_{2}-299}-2.2\times10^2 C_{2}}\right)\times 10^8 \quad \sim \mu\text{m}^2/s
\label{pa:Dl}
\end{align}

\begin{align}
\gamma_\text{eff}=\epsilon_\text{l}^{1.5}\gamma
\label{pa:Keff}
\end{align}

\begin{align}
D_\text{eff}=\epsilon_\text{l}^{0.5}D
\label{pa:Deff}
\end{align}
\begin{align}
\sigma_\text{eff}=\epsilon_\text{s}\sigma_\text{s}
\label{pa:sigmaeff}
\end{align}

\subsection{Finite strain mechanics and the evolving porosity model}
\label{sec:porosity}
Lithium intercalation/de-intercalation induces particle swelling/contraction which manifests itself as electrode deformation at the macro-scale. Additionally, the particle and separator undergo thermal expansion.\footnote{\color{black}The electrolyte is assumed not to undergo thermal expansion. Therefore the decomposition of the deformation into elastic, swelling and thermal contributions does not apply to it.} The kinematics of finite strain leads to the following decomposition:
\begin{align}
\bF=\bF^\text{e}\bF^\text{c}\bF^{\theta}
\label{eq:cellDeformGradient}
\end{align}
where $\bF = \bone + \partial\bu/\partial\bX$, is the total deformation gradient tensor averaged over the constituents of the material. It is multiplicatively decomposed into $\bF^\text{e}$, $\bF^\text{c}$ and $\bF^{\theta}$,  which are, respectively, its elastic, chemical (induced by lithium intercalation) and thermal components. In the absence of a body force the strong form of the mechanics problem in the current configuration is 
\begin{align}
\nabla\cdot\bT = \bzero
\label{eq:strongformElasticity}
\end{align}
\begin{align}
\bT= \frac{1}{\det{\bF^\text{e}}}\frac{\partial W}{\partial \bF^\text{e}}\bF^{\text{e}^\text{T}}
\end{align}
where $\bT$  is the Cauchy stress tensor and $W$ is the strain energy density function.

\begin{figure}[hbtp]
\hspace*{-1.8cm}  
\includegraphics[scale=0.4]{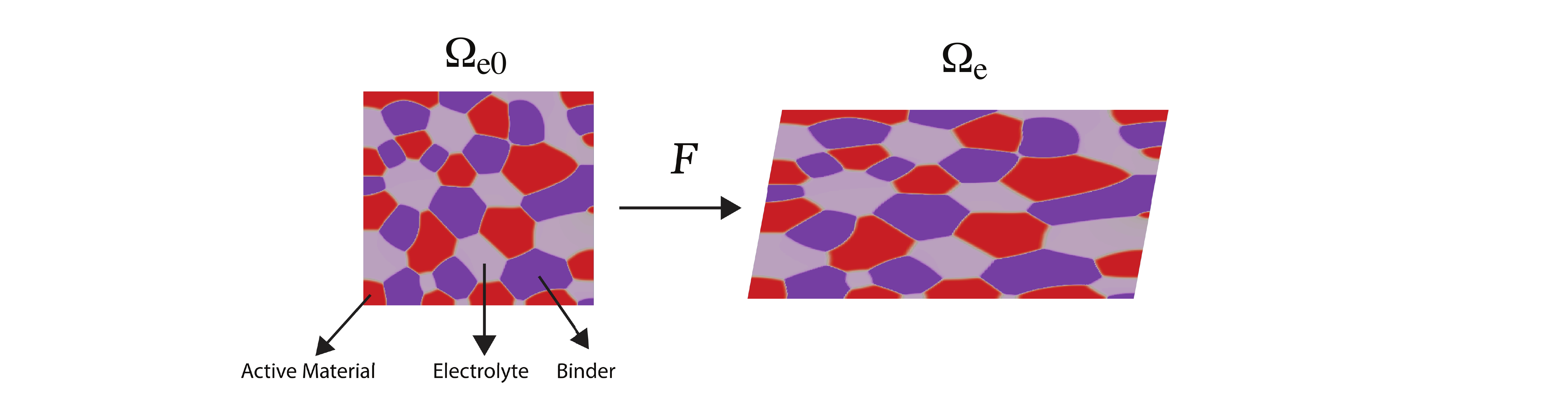}
\caption{The porous electrode mapped by $\bF$ from its reference configuration, $\Omega_{\text{e}_0}$ to its deformed configuration, $\Omega_\text{e}$.}
\label{fig:porousRVE}
\end{figure}

The electrodes are composed of solid particles, electrolyte-filled pores and binders as illustrated in Figure \ref{fig:porousRVE}. 
The sum of  volume fractions gives
\begin{align}
\epsilon_{\text{s}_0}+\epsilon_{\text{l}_0}+\epsilon_{\text{b}_0}=1  \label{eq:volumefractionInitial}\\
\epsilon_\text{s}+\epsilon_\text{l}+\epsilon_\text{b}=1\label{eq:volumefraction}
\end{align}
where $\epsilon_\text{s}$, $\epsilon_\text{l}$ have been introduced before as the volume fractions of solid particles and pores  (assumed filled with electrolyte), and  $\epsilon_\text{b}$ is the binder. The subscript $(\bullet)_0$ denotes the corresponding volume fraction in the reference configuration. For a volume element, we denote its total volume by $\delta V$ and the volume of solid particles in it by $\delta V_\text{s}$. Then, we have:
\begin{align}
\epsilon_\text{s}=\frac{\delta V_\text{s}}{\delta V}=\frac{\delta V_\text{s}}{\delta V_{\text{s}_0}}\frac{\delta V_{0}}{\delta V}\frac{\delta V_{\text{s}_0}}{\delta V_0}\\
=\frac{\det{\bF_\text{s}}}{\det{\bF}}\epsilon_{\text{s}_0} 
\label{currentsolvolfrac}
\end{align}
where $\bF_\text{s}$ is the deformation gradient tensor over the particle, $\text{det}$ denotes the determinant, and we have used $\text{det}\bF = \delta V/\delta V_0$, $\text{det}\bF_\text{s} = \delta V_\text{s}/\delta V_{\text{s}_0}$, and $\epsilon_{\text{s}_0} = \delta V_{\text{s}_0}/\delta V_0$. Similarly, we have
\begin{align}
\epsilon_\text{l}=\frac{\det{\bF_\text{l}}}{\det{\bF}} \epsilon_{\text{l}_0} \label{currentVoidVolFracFinite}\\
\epsilon_\text{b}=\frac{\det{\bF_\text{b}}}{\det{\bF}}\epsilon_{\text{b}_0} \label{currentVoidVolFracFiniteB}
\end{align}
where $\bF_\text{l}$ and  $\bF_\text{b}$ are, respectively, the deformation gradient tensors of the pore (assumed filled with electrolyte) and the binder material.
From the theory of mixtures\cite{Garikipatia2004Mixturetheory} the total stress is,
\begin{align}
\bT=\bT_\text{l}+\bT_\text{s}+\bT_\text{b}
\label{eq:MixtureStress}
\end{align}
{\color{black} We assume that the electrolyte is an ideal, incompressible fluid, implying that $\bT_\text{l}=P_\text{l}\bone$, uniform in space and constant in time. Furthermore, the stress in the binder is modelled as isotropic, uniform in space and constant in time so that $\bT_\text{b}=P_\text{b}\bone$. This implies that the elastic deformation of the binder $\bF^\text{e}_\text{b}$ is also uniform in space and constant in time. This represents a strong assumption for the binder, which we will seek to relax in future communications by taking account of the microstructure. For the electrolyte, isotropy of stress is reasonable, while its uniformity and constancy merit closer examination in microstructurally based treatments.} 

Assuming a strain energy density function $W(\bF^\text{e})$ with volumetric term 
\begin{equation}
W_\text{vol}(\bF^\text{e}) = \frac{1}{2}\kappa(\text{det}\bF^\text{e} -1)^2
\end{equation}
and the same form for the solid particle, this leads to
\begin{align}
\text{tr}\bT=\kappa(\det \bF^\text{e}-1)\text{tr}\bone, \label{eq:trStressE}\\
    \text{tr}\bT_\text{s}=\kappa_\text{s}(\det \bF^\text{e}_\text{s}-1)\text{tr}\bone. \label{eq:trStressS}
\end{align}
Computing the trace of Equation (\ref{eq:MixtureStress}), using (\ref{eq:trStressE}) and (\ref{eq:trStressS}), we have
\begin{align}
\kappa(\det \bF^\text{e}-1)=\kappa_\text{s}(\det \bF^\text{e}_\text{s}-1)+ 3P_\text{l}+ 3P_\text{b}
\label{eq:trMixtureStressFinite}
\end{align}
We write the swelling of the electrode's volume element due to lithium intercalation and thermal expansion using empirically derived functions, $\beta$ and $\beta^\theta$:
\begin{subequations}
\begin{align}
\det\bF^\text{c}&=1+\beta(C_1)\label{Fc}\\
\det\bF^\theta&=1+\beta^\theta(\theta)\label{Ftheta},
\end{align}
and similarly for its solid component:
\begin{align}
\det\bF^\text{c}_\text{s}&=1+\beta_\text{s}(C_1)\label{Fcs}\\
\det\bF^\theta_\text{s}&=1+\beta^\theta_\text{s}(\theta)\label{Fthetas}.
\end{align}
\end{subequations}
Then, the volumetric deformation of the element and its solid component can be expressed as
\begin{align}
\det \bF=\det \bF^\text{e}\det\bF^\text{c}\det\bF^\theta=\det \bF^\text{e}(1+\beta)(1+\beta^\theta)
\label{eq:Ftotal}
\end{align}
\begin{align}
\det \bF_\text{s}=\det \bF^\text{e}_\text{s}\det\bF_\text{s}^\text{c}\det\bF_\text{s}^\theta=\det \bF_\text{s}^\text{e}(1+\beta_\text{s})(1+\beta_\text{s}^\theta)
\label{eq:Fparticle}
\end{align}
{\color{black}Note that the models in Equations (\ref{Fc}--\ref{Fthetas}) represent a simplified approach as an alternative to rigorous homogenization. The lithium intercation-induced swelling of the solid component is due only to the active material, and not the binder. The effective swelling, applicable to the entire solid component has been represented by $\beta_\text{s}(C_1)$. {\color{black}Thermal expansion also is assumed to occur only in the active material, but not in the binder}; in this case, $\beta_\text{s}^\theta$ represents the effective thermal expansion over the solid component, since we do not consider the temperature field $\theta$ to vary over the spatial scale of active material and binder particles\footnote{\textcolor{black}{This assumption is valid for the through plane temperature distribution due to the relatively thin electrode structure, but should be re-assessed for the case of large format cells which could have significant in-plane temperature distributions.}}. The functions, $\beta(C_1)$ and  $\beta_\text{s}(C_1)$ have been fit to experimental data in this work (see Section \ref{sec:calibration}). The more sophisticated treatment for these functions, using analytic or computational homogenization over the microstructure, will be presented in a future communication.}

On substituting Equations (\ref{Fc}--\ref{Fthetas}) in Equation (\ref{currentsolvolfrac}) we obtain
\begin{align}
\epsilon_\text{s}=\frac{\left( \frac{\kappa(\frac{\det\bF}{(1+\beta)(1+\beta^\theta)}-1)-3P_\text{l}-3P_\text{b}}{\kappa_\text{s}} +1\right)(1+\beta_\text{s})(1+\beta_\text{s}^\theta)}{\det\bF}\epsilon_{\text{s}_0}
\label{voumefractionSolid}
\end{align}
Assuming the binder to deform at constant volume during charging and discharging such that $\text{det}\bF_\text{b}=1$, we have
\begin{align}
\epsilon_\text{b}=\frac{1}{\det \bF}\epsilon_{\text{b}_0}\label{voumefractionBinder}\\
\epsilon_\text{l}=1-\epsilon_{\text{s}}-\epsilon_\text{b}
\label{voumefractionPore}
\end{align}
 For the thermal expansion functions, we have chosen
\begin{align}
\beta^{\theta}_\text{s}(\theta)&=\Omega_\text{s} (\theta-\theta_0)\\
\beta^{\theta}(\theta)&=\Omega (\theta-\theta_0)
\end{align}
for constants $\Omega_\text{s}$ and $\Omega$.

{\color{black} Returning to Equation (\ref{eq:cellDeformGradient}), we assume that although the {\color{black} particles undergo isotropic swelling in the electrode due to intercalation of lithium, there is} unidirectional swelling along the normal to the separator; i.e., the $\be_2$-direction in Figure \ref{fig:solvingDomain}. Accordingly, we have
\begin{align}
F^\text{c}_{iJ}=\beta\delta_{2i}\delta_{2J}+\delta_{iJ} \label{eq:deformFc}
\end{align}
This assumption models the non-slip boundary condition that would be applied at the current collector-electrode interface, and which would provide a strong constraint against macroscopic expansion of the electrode in the $\be_1$ and $\be_3$ directions. Since we do not directly model the current collector, this assumption represents its mechanical effect.  On a practical note for this work, $\beta(C_1)$ and $\beta_\text{s}(C_1)$ were fit to the observed expansion along $\be_2$ and used to obtain the numerical results of Section \ref{sec:numericalexamples}. Thermal expansion is modelled as isotropic\cite{Xiao2011JPS}\footnote{\textcolor{black}{The aspect ratio of the computational domain has been chosen  with the $\be_2$ dimension to be much smaller than the other two dimensions. Additionally, the boundary conditions do not allow displacement in the $\be_1$ and $\be_3$ directions on the surfaces of the computational domain that are perpendicular to $\be_2$ (see Section \ref{sec:ibvp}). The solution to the mechanics problem therefore renders the resulting total strain (driven by intercalation and thermal expansion) to also be primarily along the $\be_2$ direction.}}. 
\begin{align}
F^{\theta}_{iJ}=(1+\beta^{\theta})^{1/3}\delta_{iJ} \label{eq:deformFT}
\end{align}}
Equations ({\ref{eq:cellDeformGradient}), ({\ref{eq:deformFc}), ({\ref{eq:deformFT}) and ({\ref{eq:strongformElasticity}) allow the volume fractions to be calculated by  also using Equations (\ref{voumefractionSolid}) and (\ref{voumefractionPore}). 

{\color{black}The separator can be treated as a medium in which the active material and binder are replaced by a porous polymer that does not undergo lithium intercalation-induced swelling, but does experience thermal expansion. Accordingly, the functions $\beta$ and $\beta_\text{s}$ vanish in the separator. The deformation of the solid component is $\text{det}(\bF_\text{s}) = \text{det}(\bF^\text{e}_\text{s})\text{det}(\bF^\theta_\text{s})$, which replaces Equation (\ref{eq:Fparticle}). The stress is $\bT = \bT_\text{l} + \bT_\text{s}$, replacing Equation (\ref{eq:MixtureStress}). The volume fraction expressions in the separator reduce to
\begin{align}
\epsilon_\text{s}=\frac{\left( \frac{\kappa(\frac{\det\bF}{1+\beta^\theta}-1)-3P_\text{l}}{\kappa_\text{s}} +1\right)(1+\beta_\text{s}^\theta)}{\det\bF}\epsilon_{\text{s}_0}
\label{voumefractionSolidSep}
\end{align}
and since the binder is absent,
\begin{equation}
\epsilon_\text{l}=1-\epsilon_{\text{s}}.
\label{volumefractionSep}
\end{equation}
}

\subsection{Revised specific area and $C^{\text{max}}_1$}
\label{sec:revisedCmax}
{\color{black}Since we assume that lithium intercalation makes the spherical particles swell isotropically, the particle radius needs to be updated as follows to determine the specific area, $a_\text{p}=S_\text{p}/V_\text{p} = 3/R_\text{p}$. Here,} the particle volume is $V_\text{p}=V_{\text{p}_0}\det{\bF_\text{s}}$. From equations (\ref{eq:trMixtureStressFinite}), (\ref{eq:Ftotal}) and (\ref{eq:Fparticle}), this gives
\begin{align}
V_\text{p}=V_{\text{p}_0}\left(\frac{\left(\kappa(\frac{\det\bF}{(1+\beta)(1+\beta^\theta)}-1)-3P_\text{l}-3P_\text{b}\right)}{\kappa_\text{s}} +1\right)(1+\beta_\text{s})(1+\beta_\text{s}^\theta),
\label{eq:vpvp0}
\end{align}
from which the spherical radius can also be written as 
 \begin{align}
 R_\text{p}=\left(\frac{3}{4\pi}V_\text{p}\right)^{\frac{1}{3}} \label{eq:Rp}
 \end{align} 
As the particles swell/contract due to lithium intercalation/de-intercalation, the maximum concentration of lithium that they can accommodate varies. We define $C^{\text{max}_0}_1$ to be the maximum concentration of lithium in the undeformed particle. This quantity is a material constant. Furthermore, as the particle swells/contracts, the maximum number of lithium atoms that it can accommodate remains fixed. This implies that
 \begin{align}
 C^{\text{max}}_1V_\text{p}=C^{\text{max}_0}_1V_{\text{p}_0}
 \end{align}
 from which, using Equation (\ref{eq:vpvp0}), $C^{\text{max}}_1$ is given by
 \begin{align}
 C^{\text{max}}_1=\frac{C^{\text{max}_0}_1}{\left[\left(\frac{\left(\kappa(\frac{\det\bF}{(1+\beta)(1+\beta^\theta)}-1)-3P_\text{l}-3P_\text{b}\right)}{\kappa_\text{s}} +1\right)(1+\beta_\text{s})(1+\beta_\text{s}^\theta)\right]}
 \label{eq:Cmax}
 \end{align}
\subsection{The analytic diffusion profile for lithium in a particle}
\label{sec:parabolicprofile}
We note that Equation (\ref{eq:conserC1Strong}) is derived by using the particle volume-averaged concentration, $C_1$, but ignoring diffusion of lithium in the solid particle. It is traditional to assume a spherically symmetric distribution of lithium within the particle due to uniform boundary conditions on the particle surface, with a parabolic profile along the particle radius\cite{Smith2006}. Letting $c_1$ denote the intra-particle lithium concentration, its relation to the volume-averaged concentration, $C_1$, is

\begin{align}
C_{1_\text{surf}}=c_1 \vert_{r=R_\text{p}}=C_1-\frac{R_\text{p}j_n}{5D_\text{s}}.
\label{eq:C1surface}
\end{align}
{\color{black} Where $R_\text{p}$ is the time varying particle radius due to swelling given by Equation (\ref{eq:Rp}).}
The detailed derivation appears in the Appendix.


\section{Numerical treatment}
\label{sec:numerics}
Equations (\ref{eq:conserC1Strong}), (\ref{eq:conserC2Strong}), (\ref{eq:phieEq}-\ref{eq:ThermalSeparator}) and (\ref{eq:strongformElasticity}) are coupled and highly nonlinear. Furthermore, the coefficients in the partial differential equations and many response functions, (\ref{pa:irreversibleentropicheat}--\ref{pa:sigmaeff}), (\ref{voumefractionSolid}--\ref{eq:deformFT}) and (\ref{eq:Cmax}) depend on the primary variables, introducing further nonlinearity to the system of equations. Here, they are written in weak form and solved by the finite element method using code implemented in the open source finite element library {\tt deal.II}\cite{Bangerth2007,dealII84}.

\subsection{The Galerkin weak form and the finite element formulation}
For a generic, finite-dimensional field $u^h$, the problem is stated as follows: Find $u^h\in \mathscr{S}^h \subset \mathscr{S}$, where $\mathscr{S}^h= \{ u^h \in \mathscr{H}^1(\Omega_0) ~\vert  ~u^h = ~\bar{u}\; \mathrm{on}\;  \Gamma_{0}^u\}$,  such that $\forall ~w^h \in \mathscr{V}^h \subset \mathscr{V}$, where $\mathscr{V}^h= \{ w^h \in\mathscr{H}^1(\Omega_0)~\vert  ~w^h = ~0 \;\mathrm{on}\;  \Gamma_{0}^u\}$, the finite-dimensional (Galerkin) weak form of the problem is satisfied. The variations, $w^h$ and trial solutions $u^h$ are defined component-wise using a finite number of basis functions,
\begin{equation}
w^h = \sum_{a=1}^{n_\mathrm{b}} c^a N^a, \quad \qquad u^h = \sum_{a=1}^{n_\mathrm{b}} d^a N^a 
\label{eq:basisdef}
\end{equation}
\noindent where $n_\mathrm{b}$ is the dimensionality of the function spaces $\mathscr{S}^h$ and $\mathscr{V}^h$, and $N^a$ represents the basis functions.

To obtain the Galerkin weak form for each strong form (\ref{eq:conserC1Strong}), (\ref{eq:conserC2Strong}), (\ref{eq:phieEq}-\ref{eq:ThermalSeparator}) and (\ref{eq:strongformElasticity}), we multiply the strong form by the corresponding weighting function, integrate by parts and apply boundary conditions appropriately. The final weak form expressed in terms of residual equations appears below. 
\begin{equation}
\mathscr{R}_{C_1}=\int_{\Omega_{\text{e}}}w_{c_1} \epsilon_\text{s}\frac{\partial C_1}{\partial t}\text{d}v+\int_{\Omega_{\text{e}}}w_{c_1} C_1\frac{\partial \epsilon_\text{s}}{\partial t}\text{d}v + \int_{\Omega_{\text{e}}}w_{c_1}\epsilon_\text{s}a_\text{p}{j_{n}}\text{d}v = \bzero,
\label{eq:WeakformR1}
\end{equation}

\begin{align}
\mathscr{R}_{C_2}=\int_{\Omega_{\text{e}}}w_{c_2} \epsilon_\text{l}\frac{\partial C_2}{\partial t}\text{d}v+\int_{\Omega_{\text{e}}}w_{c_2} C_2 \frac{\partial \epsilon_\text{l}}{\partial t}\text{d}v -\int_{\Omega_{\text{e}}}w_{c_2}(1-t^{0}_{+})\epsilon_\text{s}a_\text{p}j_{n}\text{d}v\nonumber\\
+\int_{\Omega_{\text{e}}}\nabla w_{c_2} D_\text{eff}\nabla C_{2}\text{d}v-\int_{S}w_{c_2} D_\text{eff}\nabla C_{2}\cdot \bn \text{d}S = \bzero,
\label{eq:WeakformR2}
\end{align}
\begin{equation}
\mathscr{R}_{\phi_\text{S}}=-\int_{\Omega_{\text{e}}}\nabla w_{\phi_\text{S}}[\sigma_\text{eff}(-\nabla\phi_\text{S})]\text{d}v+\int_{\Omega_{\text{e}}}w_{\phi_\text{S}} aFj_{n}\text{d}v+\int_{S}w_{\phi_\text{S}}[\sigma_\text{s}(-\nabla\phi_\text{S})]\cdot \bn \text{d}S = \bzero,
\label{eq:WeakformR3}
\end{equation}
\begin{align}
\mathscr{R}_{\phi_\text{E}}=-\int_{\Omega_{\text{e}}}\nabla w_{\phi_\text{E}}[\gamma_\text{eff}(-\nabla\phi_\text{E})+\gamma_\text{eff}\frac{2RT}{F}(1-t^{0}_{+})\nabla\ln C_2]\text{d}v-\int_{\Omega_{\text{e}}}w_{\phi_\text{E}} a_\text{p}Fj_{n}\text{d}v\nonumber
\\
+\int_{S}w_{\phi_\text{E}}[\gamma_\text{eff}(-\nabla\phi_\text{E})+\gamma_\text{eff}\frac{2RT}{F}(1-t^{0}_{+})\nabla\ln C_2]\cdot \bn \text{d}S = \bzero,
\label{eq:WeakformR4}
\end{align}
\begin{align}
\mathscr{R}_\theta=\int_{\Omega_{\text{e}}}w_\theta\rho C_p \frac{\partial \theta}{\partial t}\text{d}v+\int_{\Omega_{\text{e}}}\nabla w_\theta \lambda \nabla \text{d}v- \int_{\Omega_{\text{e}}} w_\theta Q  \text{d}v- \int_{S}w_\theta\lambda \nabla \theta\cdot \bn \text{d}S = \bzero,
\label{eq:WeakformR5}
\end{align}
\begin{align}
\mathscr{R}_u=\int_{\Omega_{\text{e}}}\nabla \bw_u\bT \text{d}v- \int_{S}\bw_u \bff \cdot \bn \text{d}S = \bzero,
\label{eq:WeakformR6}
\end{align}
where $Q=Q_{\text{rxn}}+Q_{\text{rev}}+Q_{\text{ohm}}$ in the electrodes as in Equation (\ref{eq:ThermalElectrode}) and $Q=Q_{\text{ohm}}$ in the separator as in Equation (\ref{eq:ThermalSeparator}). The fields $w_{c_1}$, $w_{c_2}$, $w_{\phi_\text{S}}$, $w_{\phi_\text{E}}$, $w_\theta$ and $\bw_u$ are weighting functions for the corresponding primary variables. Time integration is achieved by the Backward-Euler algorithm.}

\subsection{Algorithmic differentiation}
The analytical linearization of the residual equations (\ref{eq:WeakformR1}-\ref{eq:WeakformR6}) to obtain the Jacobian matrix is tedious and is fraught with the danger of algebraic mistakes. Symbolic differentiation is an option, but its speed can be a limitation for complicated nonlinear and coupled problems such as those in the present communication. An easy alternative is the use of numerical differentiation tools built into many standard solver packages. However, for a highly non-linear set of equations, numerical differentiation is inaccurate and ultimately unstable. An effective and efficient alternative is the use of algorithmic (or automatic) differentiation (AD), which works by application of the chain rule to algebraic operations and functions (polynomial, trigonometric, logarithmic, exponential or reciprocal) in the code. AD thus works to machine precision at a computational cost that is comparable to the cost of evaluation of the original equations. We use AD in this work to linearize Equations (\ref{eq:WeakformR1}-\ref{eq:WeakformR6}), and compute the Jacobian matrix. Specifically, we use the {\tt Sacado} package, which is part of the open-source {\tt Trilinos} project\cite{Sacado2005,sacado2012}.

\section{Numerical examples}
\label{sec:numericalexamples}
Using a prototype Li ion battery cell, we present a number of numerical results to demonstrate (\romannumeral 1) the evolution of porosity  driven by lithium intercalation/de-intercalation, thermal expansion and mechanical stresses, and (\romannumeral 2) the influence of the evolving porosity, via electrostatic and reaction-transport coefficients, on the battery's characteristics. We present as a benchmark, a computation with fixed porosity, i.e. decoupled from the effects of lithiation and strain, and under isothermal conditions. It is compared with other computations that include non-uniform lithiation, non-isothermal conditions and thermal strain effects.
\subsection{The initial/boundary value problem}
\label{sec:ibvp}
We model a three dimensional cell (Figure \ref{fig:solvingDomain}) of size $120,000\mu\text{m}\times 128\mu\text{m}\times85,000\mu\text{m}$. For ease of interpretation of the coupled physics in this first communication, the boundary conditions and distribution of coefficients have been chosen to render the problems to be effectively one-dimensional.  { \color{black} The fields do not vary along the $\be_1$ and $\be_3$ directions. Therefore a single element sufficed along these directions. An element width of 1$\mu\text{m}$ was used along the $\be_2$ direction. Trilinear hexahedral elements were used. Because of the one-dimensional nature of the problems considered, the high element aspect ratios did not manifest in numerical ill-conditioning. In order to overcome the stiff dynamics using the Backward Euler algorithm, initial time steps of $0.1$s were used, gradually increasing to $10$s for computations at the 1~C rate, and $1$s for the 10~C rate.\footnote{A 1~C rate means that a battery rated at N Ah should supply N Amperes for 1 hour.}} Details of all parameters and initial conditions are summarized in the Appendix. The partial differential equations solved are either parabolic (Equations (\ref{eq:WeakformR2}) and (\ref{eq:WeakformR5})) or elliptic (Equations (\ref{eq:WeakformR3}), (\ref{eq:WeakformR4}) and (\ref{eq:WeakformR6})), and so boundary conditions must be specified on each surface. {\color{black}Conservation of lithium ions in the electrolyte translates to zero flux boundary conditions for $C_2$. Since $\phi_\text{S}$ is only defined over the electrodes, it has boundary conditions specified on the domain boundaries, and interface conditions at the electrode-separator interfaces. Its boundary conditions are $\phi_\text{S} = 0$ at $x_2 = 0$ (reference potential), and $-\sigma_\text{eff}\nabla\phi_\text{S}\cdot \bn=I/F$ at $x_2 = 128\mu\text{m}$ (applied current).  The boundary conditions on the field in the electrolyte are $\nabla\phi_\text{E}\cdot\bn = 0$ on all boundaries. Boundary conditions for the thermal field, $\theta$, correspond to conductive heat transfer to the ambient air which is assumed to be at $25^\circ\text{C}$. For mechanics, displacement boundary conditions are prescribed, $\bu = \bzero$ at $x_2 = 0$, and $\bu = u_0\be_2$ at $x_2 = 128\mu\text{m}$. These mechanical boundary conditions compress the cell to a fixed total strain as is common in automotive battery packs. The remaining surfaces are taken to be traction-free. The boundary and interface conditions are summarized in Table \ref{tbl:boundaryconditions}, where the surfaces referred to are depicted in Figure \ref{fig:solvingDomain}. }

\begin{table}[]
\centering
\caption{The boundary conditions applied on each of the governing partial differential equations, and electrode-separator interface conditions.}
\begin{tabular}{ccc}
 \hline
$\mathscr{R}_{C_1}$:&{\color{black}-}& {\color{black}-}\\ \hline
$\mathscr{R}_{C_2}$:& $\nabla C_2\cdot \bn=0 $& \text{on all surfaces}\\  \hline
$\mathscr{R}_{\phi_\text{S}}$:&$ -\sigma_\text{eff}\nabla\phi_\text{S}\cdot \bn=I/F $& \text{on surface 2}\\
       &$\phi_\text{S}=0$& \text{on surface 1}\\
        &$ [-\nabla\phi_\text{S})]\cdot \bn=0$& {\color{black}\text{on interfaces and remaining boundary surfaces}}\\  \hline
$\mathscr{R}_{\phi_\text{E}}$:&$[\gamma_\text{eff}(-\nabla\phi_\text{E})+\gamma_\text{eff}\frac{2R\theta}{F}(1-t^{0}_{+})\nabla\ln C_2]\cdot \bn=0$& \text{on all surfaces}\\  \hline
$\mathscr{R}_\theta$:&$-\lambda \nabla \theta\cdot \bn=h(\theta-\theta_\text{air})$& \text{on all surfaces}\\  \hline
$\mathscr{R}_u$:&$\bu=u_0\be_2 \quad \bu=0  $  & \text{on surface 1 and 2}\\
        &$\bT\cdot \bn=\mathbf{0}$& \text{on other surfaces}\\
        \hline
\end{tabular}
\label{tbl:boundaryconditions}
\end{table}

\begin{figure}[hbtp]
\hspace*{-3cm}  
\includegraphics[scale=0.4]{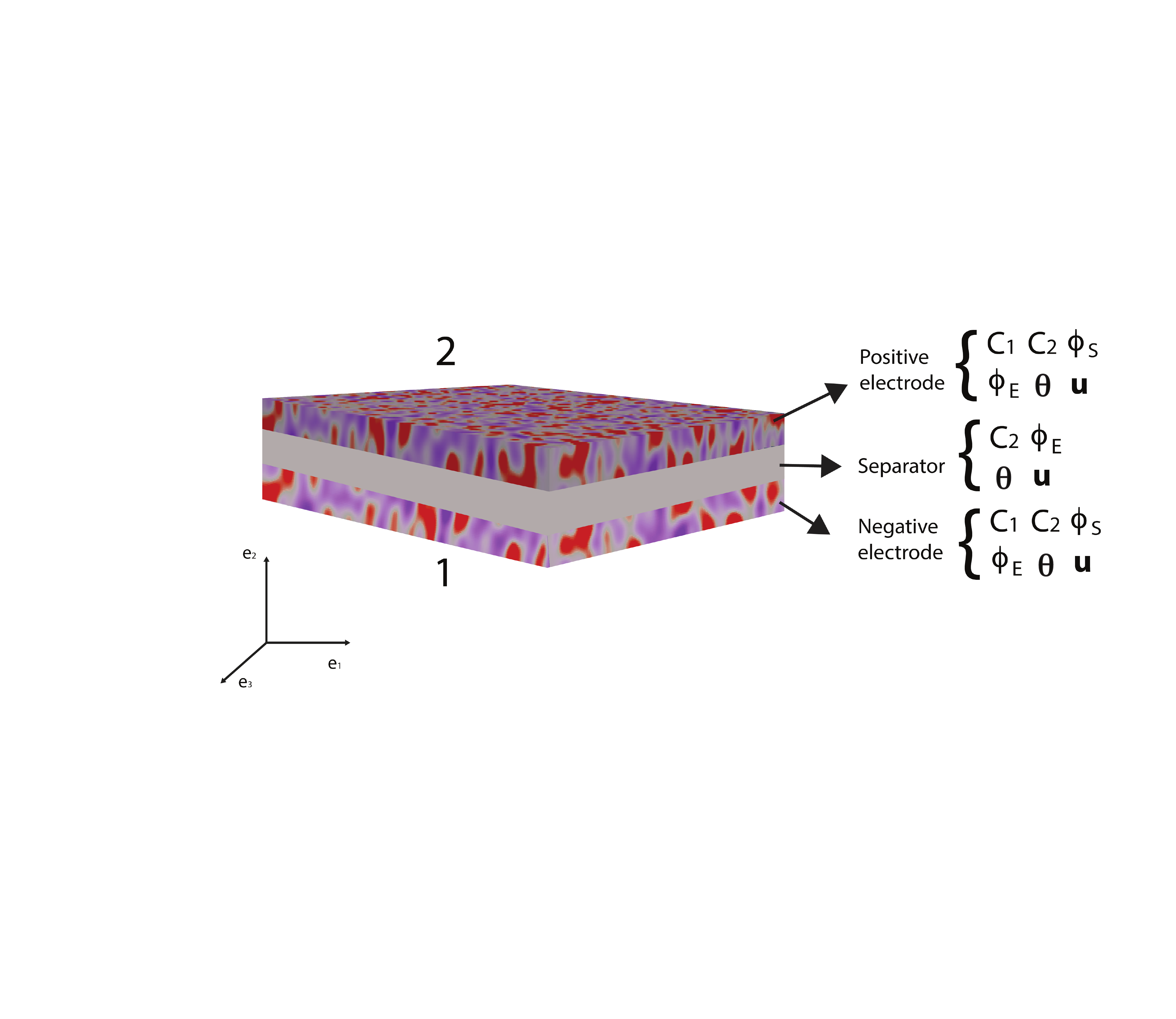}
\caption{A schematic of the initial/boundary value problem showing the fields solved for in the electrode and separator sub-domains, with the surfaces labelled.}
\label{fig:solvingDomain}
\end{figure}
We computed a full cycle (discharging and charging) under 10~C and 1~C current rates for the cases outlined in (\romannumeral 1) and (\romannumeral 2) above. For cases with decoupled porosity, the initial porosity, $\epsilon_{\text{s}_0}$, from which discharging happens, has been set higher than in the coupled porosity case to ensure that when fully discharged, both these models have the same porosity. This leads to smaller initial volume fractions of solid particles for cases with decoupled porosity. However, the total number of moles of lithium in the electrode is identical for both cases when they start at the same state of charge (SOC). The parameter $C^{\text{max}}_1$ therefore changes as expressed by Equation (\ref{eq:Cmax}). All initial porosity distributions were taken to be uniform. For studies of the effect of porosity we considered two cases: (a) $\beta_\text{s} \neq 0$, $\beta=0$; i.e., {\color{black}particle swelling is accommodated within the electrode free volume} and (b) $\beta_\text{s} \neq 0$, $\beta \neq 0$; particle swelling causes electrode swelling. {\color{black} Here, SOC was calculated by coulomb counting. Since constant current rates were applied, the SOC was defined as $\text{SOC}=1-\frac{\text{current time}}{\text{total time}}$ for discharging, or $\text{SOC}=\frac{\text{current time}}{\text{total time}}$ for charging, where the total time was for fully discharging or fully charging the cell.} {\color{black} These definitions are equivalent to a volume averaged notion of SOC, given by $\text{SOC}=(\bar{\eta} - C_{\text{1soc}_0})/(C_{\text{1soc}_{100}}-C_{\text{1soc}_0})$, where $\bar{\eta}$ is the electrode volume average of $\eta$, \textcolor{black}{recalling the definition $\eta=C_1/C_1^\text{max}$}, and $C_{\text{1soc}_y}$ is the fraction of the theoretical electrode maximum lithium content at the given SOC. Thus defined, SOC is equivalent for both electrodes in the absence of side reactions or loss of lithium. }

\subsection{Calibration of response functions}
\label{sec:calibration}
For our prototypical Li ion battery cell, the swelling function $\beta$ was fit to data of Oh et al.\cite{KiYong2014}--see Equation (\ref{eq:beta}) and Figure \ref{fig:beta}--and $\beta_\text{s}$ was fit to data of Takami et al.\cite{Takami1995}--see Equation (\ref{eq:beta_s}) and Figure \ref{fig:beta_s}. The swelling functions were redefined as, $\beta(C_1)= \hat{\beta}(\eta)$ and $\beta_\text{s}(C_1)= \hat{\beta}_\text{s}(\eta)$, with
\begin{align}
\hat{\beta}(\eta)& =\frac{0.0189\eta^5-0.039\eta^4+0.053\eta^3-0.034\eta^2+0.009\eta-0.0002}{\eta^2-0.885\eta+0.258}, \label{eq:beta} \\
\hat{\beta}_\text{s}(\eta) &=1.496\eta^3-1.739\eta^2+1.020\eta-0.033\exp(2.972\eta)-0.046\tanh(\frac{\eta-0.1}{0.1})\nonumber\\
              &-0.004\tanh(\frac{\eta-0.3}{0.1})+0.021\tanh(\frac{\eta-0.65}{0.1}).
\label{eq:beta_s}
\end{align}

\begin{figure}[hbtp]
\centering
\includegraphics[scale=0.3]{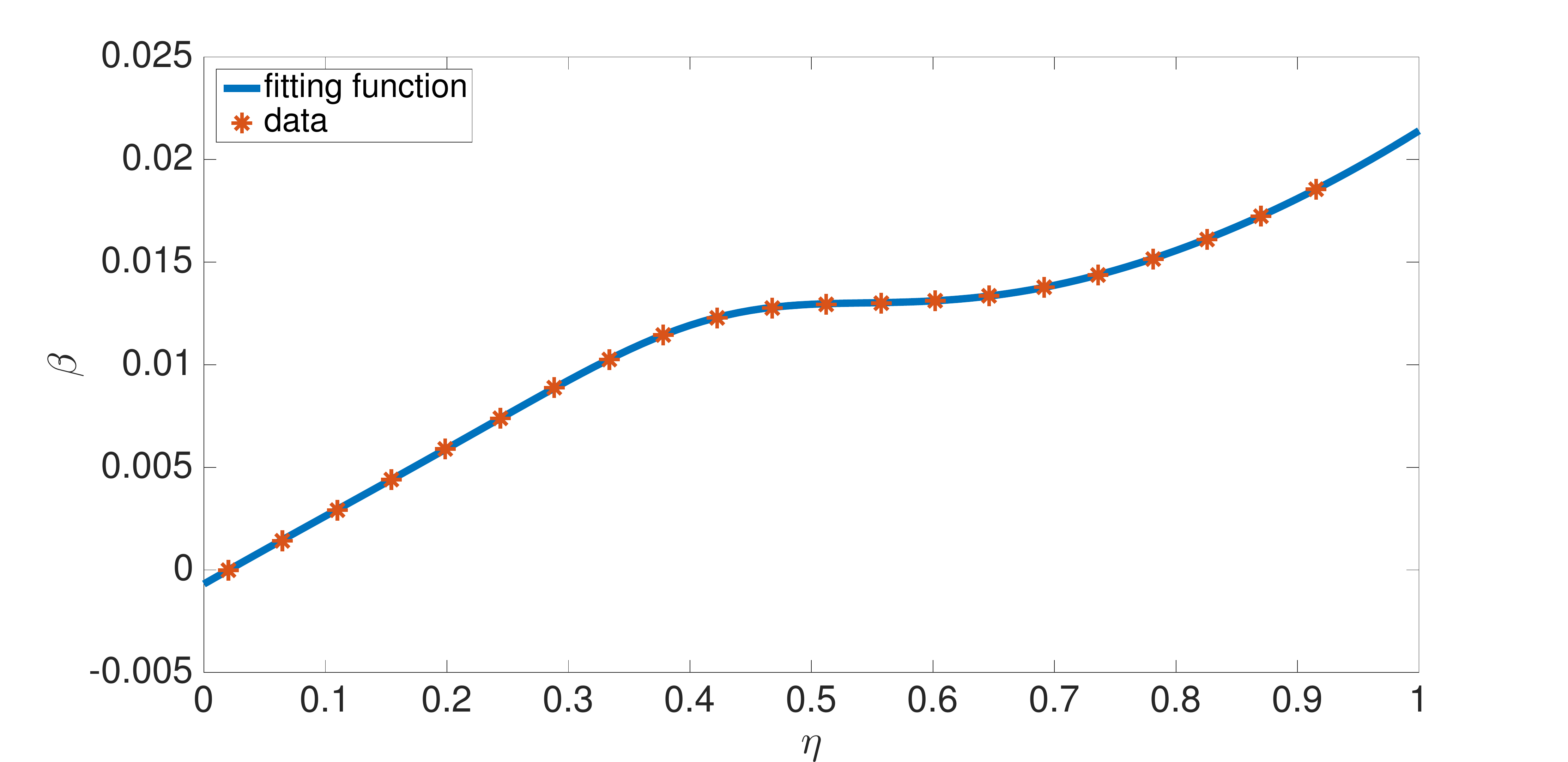}
\caption{Data for the electrode lithiation expansion function, $\beta$, is from Oh et al.\cite{KiYong2014}. The solid curve is a fit given by Equation (\ref{eq:beta}). Here, $ \eta=\frac{C_1}{C_1^\text{max}}$.}\label{fig:beta}
\end{figure}

\begin{figure}[hbtp]
\centering
\includegraphics[scale=0.3]{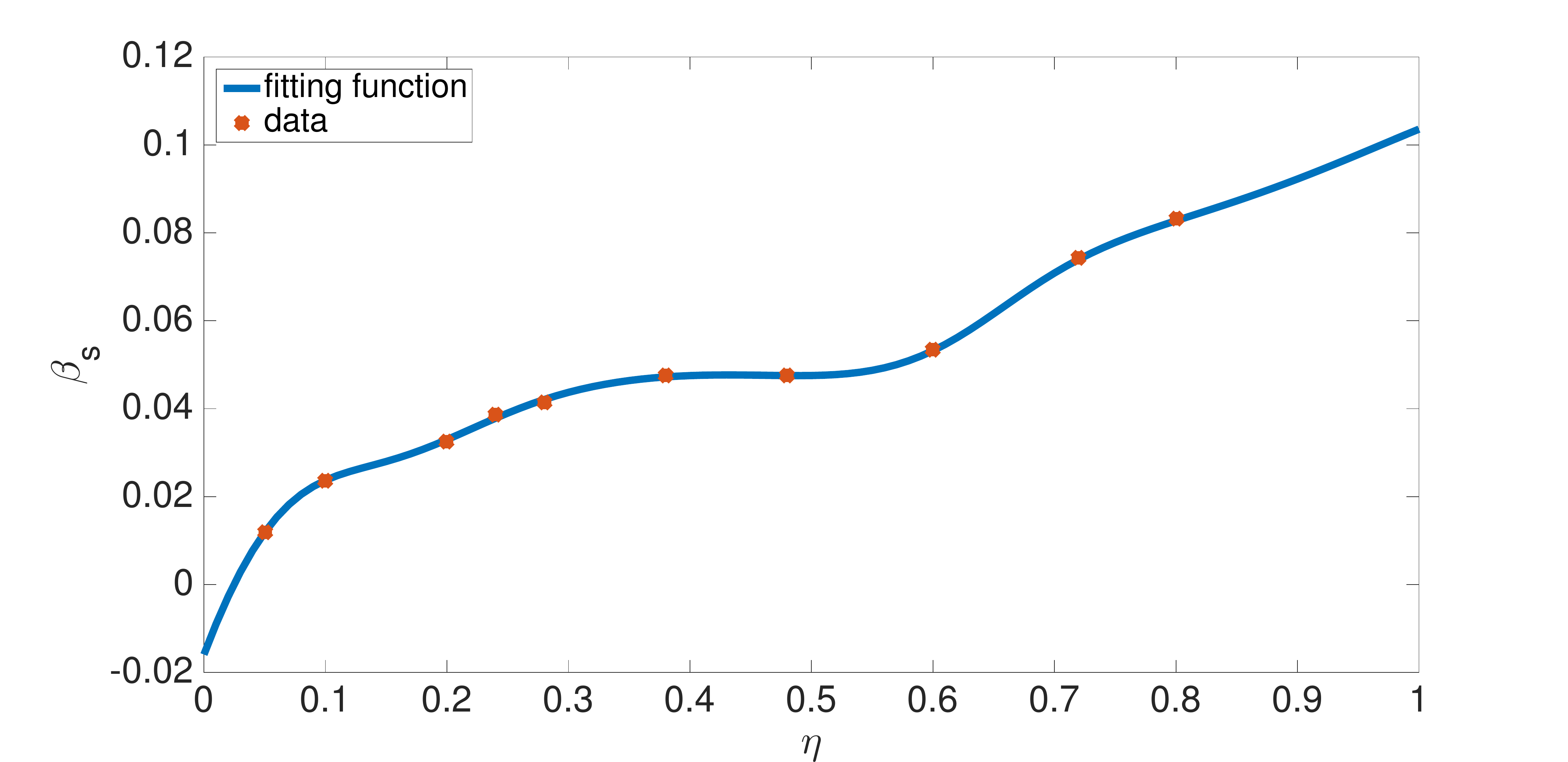}
\caption{Data for the particle expansion function, $\beta_\text{s}$, is from Takami et al.\cite{Takami1995}. The solid curve is a fit given by Equation (\ref{eq:beta_s}). Here, $ \eta=\frac{C_1}{C_1^\text{max}}$. }
\label{fig:beta_s}
\end{figure}
{\color{black}Lithium intercalation and de-intercalation were modelled only in the negative electrode. In the positive electrode we assume $\beta(C_1)=0$ and $\beta_\text{s}(C_1)=0$, motivated by experimental observations\cite{Wang2012} that the positive electrode has insignificant swelling. {\color{black}While $P_\text{l}$ and $P_\text{b}$ are not evaluated numerically in our computations, we assume that $P_\text{l} \ll \kappa_\text{s}$, and $P_\text{b} \ll \kappa_\text{s}$. Therefore, the actual values chosen for $P_\text{l}$ and $P_\text{b}$ do not influence the porosity strongly, and these parameters were set to zero in the numerical examples.}} All other parameter values are listed in the Appendix.  

\clearpage

\subsection{Computational results}
With $\beta_\text{s} = 0, \beta = 0$, there is no expansion/contraction of the particle and composite electrode due to lithium intercalation/de-intercalation. For $\beta_\text{s} \neq 0, \beta = 0$, the particle expands with Li intercalation; however, the composite electrode does not itself expand but accommodates the expanding particle in the microstructure's free volume. For $\beta_\text{s} \neq 0, \beta \neq 0$, both the particle and composite electrode expand with Li intercalation. Figures \ref{fig:10Cepl-nothermal} and \ref{fig:10Cepl-thermal}, respectively, show the porosity distribution through the thickness of the electrodes and separator (along the $x_2$ coordinate) at the indicated state of charge for the 10~C rate, under isothermal and non-isothermal (governed by the heat equation) conditions. The compressive boundary conditions cause porosity to decrease rapidly in the separator as the free volume is compressed, while the porosity decrease in the stiffer electrodes is relatively small. This is seen in the black (initial) and purple (very short times after start of the computation) curves of Figure \ref{fig:10Cepl-nothermal}. The porosity has fallen further to $\epsilon_\text{l} \sim 0.32$ by SOC = 91\% . During discharge, lithium undergoes de-intercalation from the solid particles in the negative electrode, causing its porosity to increase by contraction. When the cell is fully discharged, the porosity has increased to about $\epsilon_\text{l} = 0.35$ in the negative electrode. We note that the negative electrode also undergoes mechanical contraction due to the de-intercalation. When fully discharged, the volume of the negative electrode therefore contracts by about 9$\%$. This results in a smaller increase in porosity compared with the case of no expansion of the composite electrode ($\beta_\text{s} \neq 0, \beta = 0$) for which $\epsilon_\text{l} \sim 0.36$. {\color{black}We assume there is no intercalation {\color{black} strain} in the positive electrode as justified above (see Section \ref{sec:calibration}). Therefore its deformation is only due to stress transmitted from the negative electrode's expansion/contraction, and due to its own thermal strain. However, the small thermal expansion results in insignificant dimensional change, and since the separator is one order of magnitude less stiff than the electrodes, the stress state induced under the applied displacement boundary condition causes insignificant strain in the positive electrode.  As a result, the porosity of the positive electrode remains fairly constant. }
\begin{figure}[hbtp]
\centering
\includegraphics[scale=0.3]{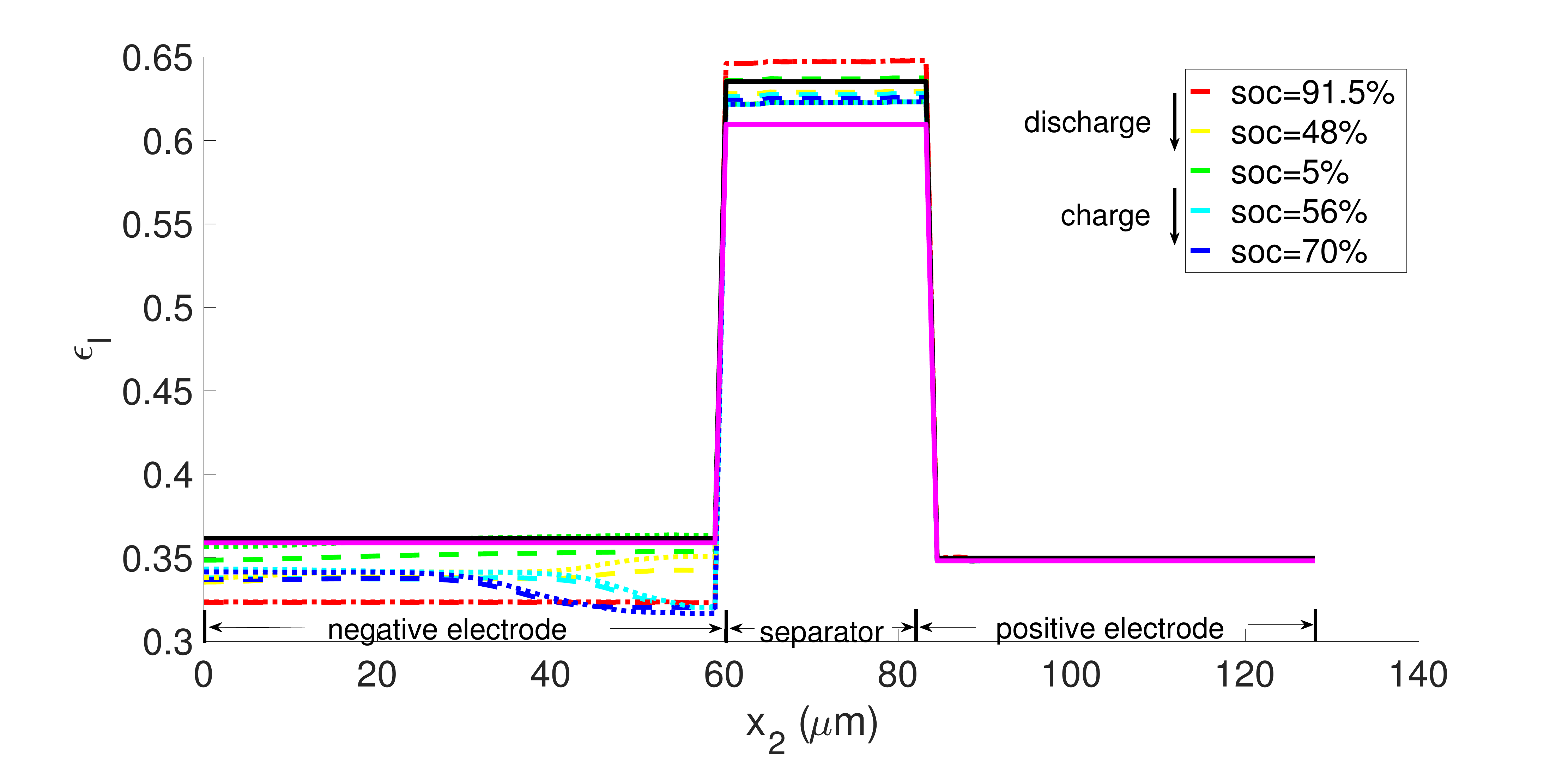}
\includegraphics[scale=0.3]{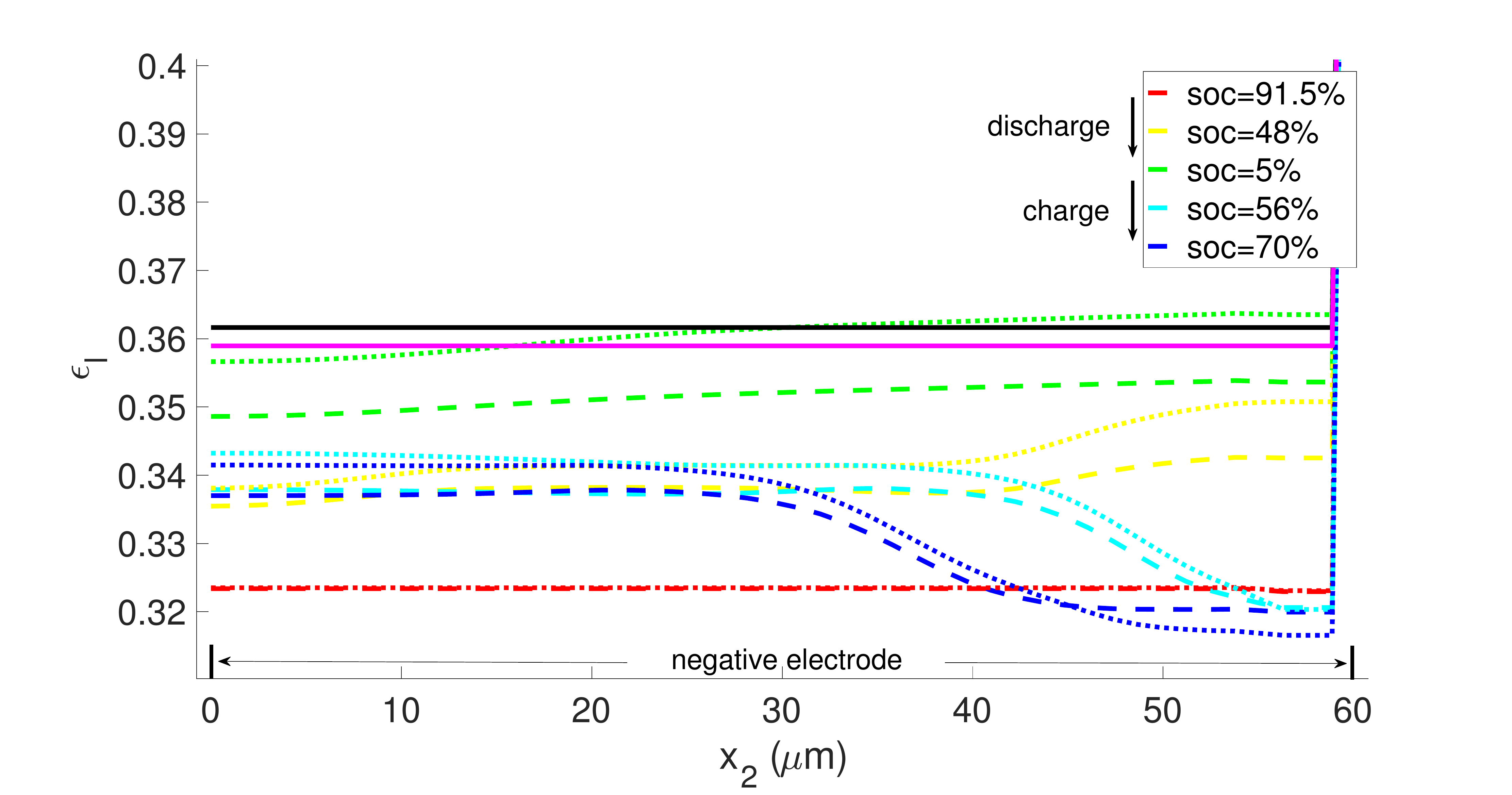}
\caption{Porosity evolving at the 10~C rate under assumed isothermal conditions; i.e., without solving the heat equation. Dashed curves: $\beta_\text{s} \neq 0$, $\beta \neq 0$;  dotted curves: $\beta_\text{s} \neq 0$, $\beta = 0$. The two solid lines represent the porosity at the initial state (black) and just after compression (purple). The porosity is highly non-uniform in the negative electrode and does not recover its initial distribution.} 
\label{fig:10Cepl-nothermal}
\end{figure}

\begin{figure}[hbtp]
\centering
\includegraphics[scale=0.3]{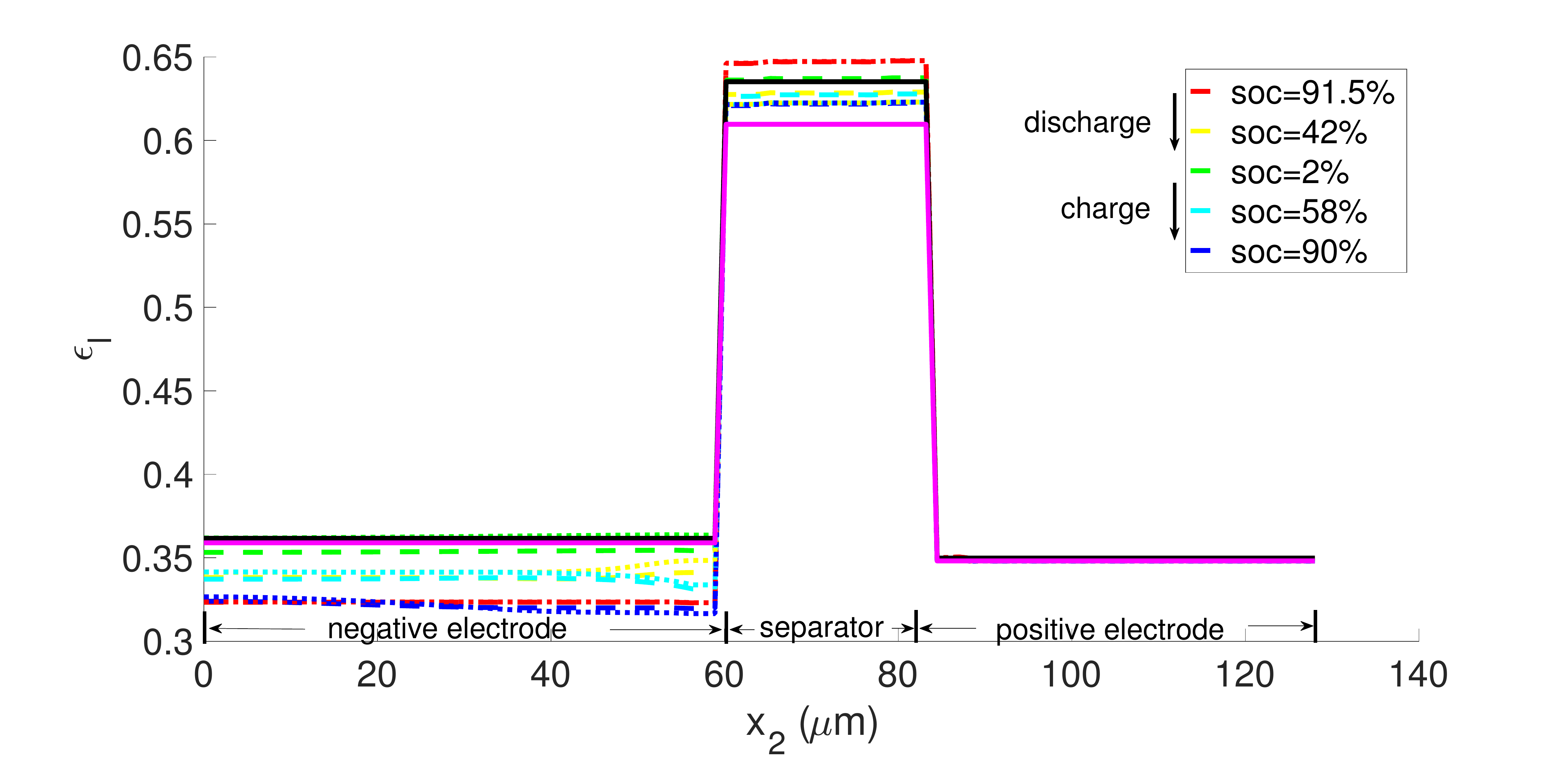}
\includegraphics[scale=0.3]{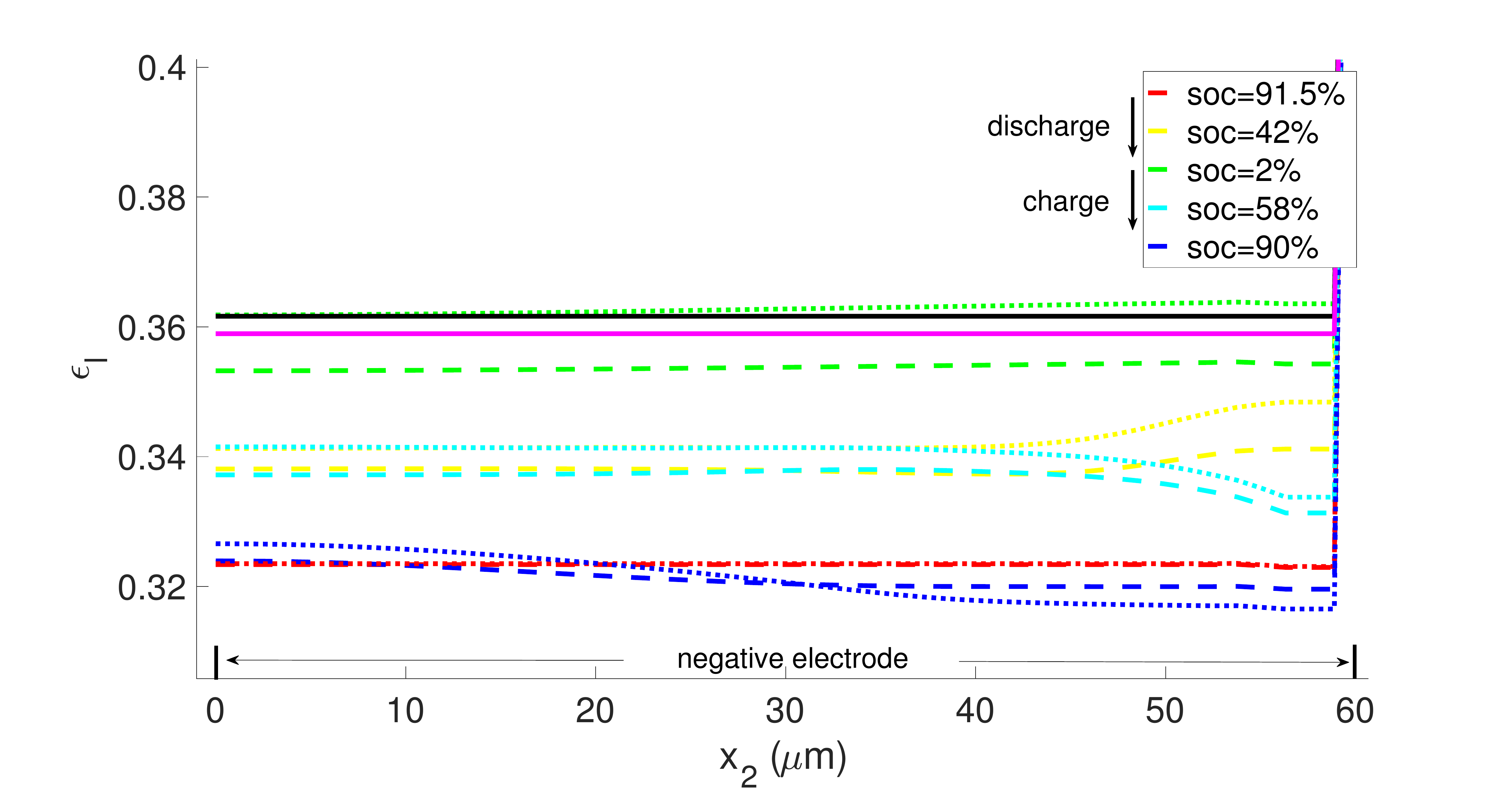}
\caption{ Porosity evolving at the 10~C rate under non-isothermal conditions. Dashed curves: $\beta_\text{s} \neq 0$, $\beta \neq 0$;  dotted curves: $\beta_\text{s} \neq 0$, $\beta = 0$. The two solid curves represent the porosity at the initial state (black) and just after compression (purple).} 
\label{fig:10Cepl-thermal}
\end{figure}

\begin{figure}[hbtp]
\centering
\includegraphics[scale=0.3]{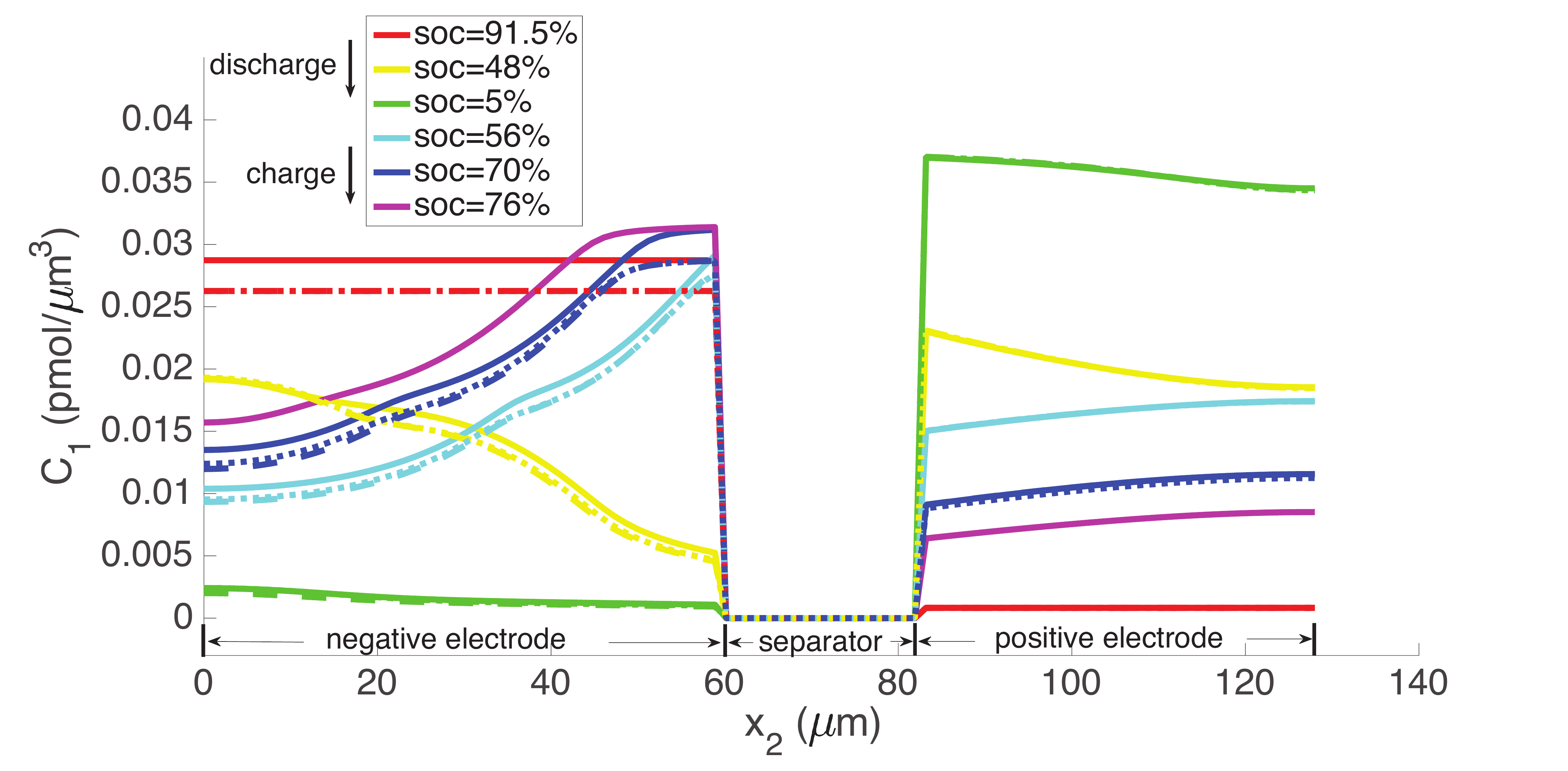}
\caption{Lithium concentration at 10~C rate under isothermal conditions. Solid curves: $\beta_\text{s}=0$, $\beta=0$; dashed curves: $\beta_\text{s} \neq 0$, $\beta \neq 0$;  dotted curves: $\beta_\text{s} \neq 0$, $\beta = 0$. Note, that for $\beta_\text{s}=0$, $\beta=0$ the battery can be recharged up to SOC = 76\% (solid purple line).} 
\label{fig:10CC1-nothermal}
\end{figure}

\begin{figure}[hbtp]
\centering
\includegraphics[scale=0.3]{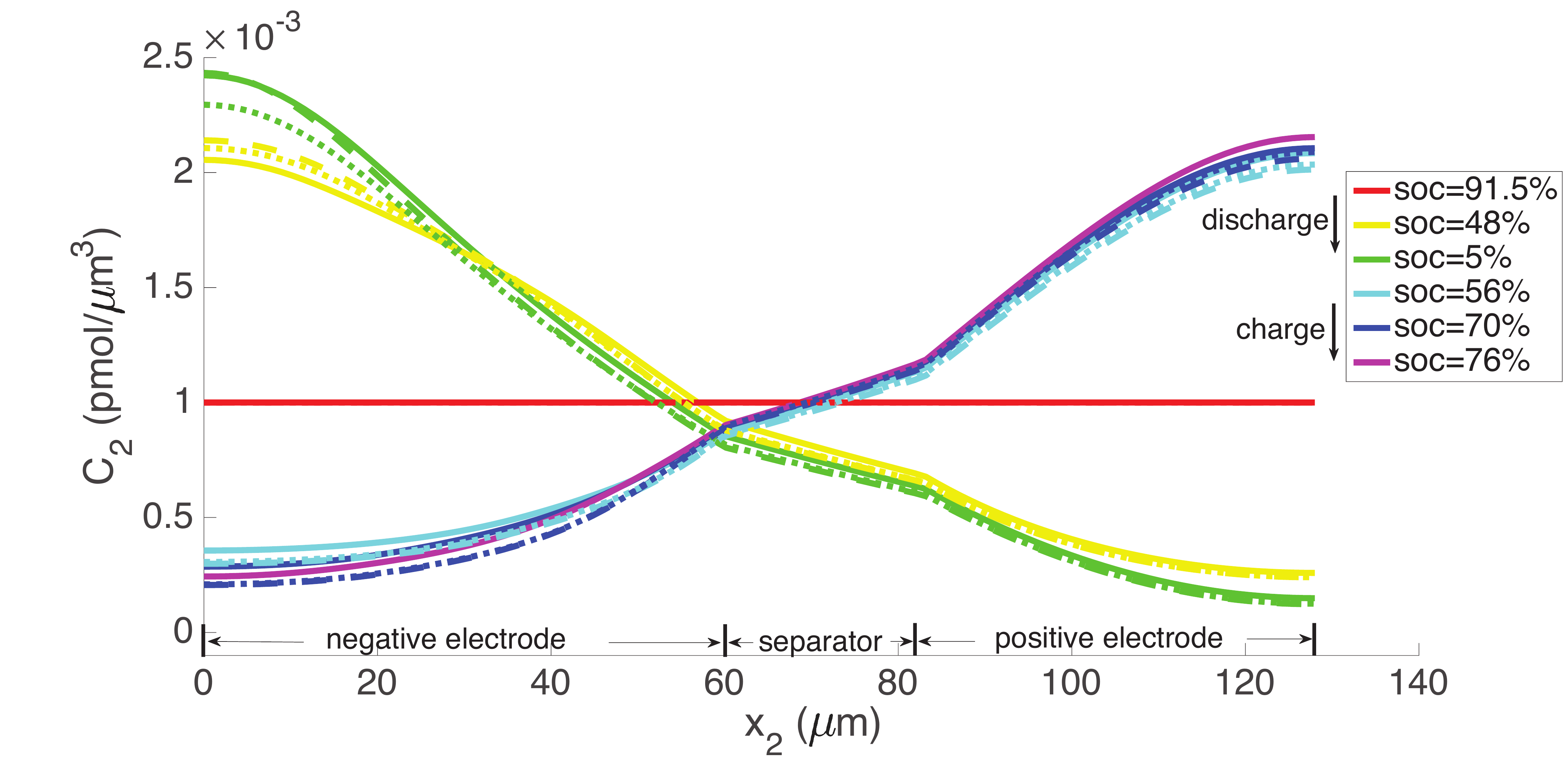}
\caption{Lithium ion concentration at a 10~C rate under isothermal conditions. Solid curves: $\beta_\text{s}=0$, $\beta=0$ ; dashed curves: $\beta_\text{s} \neq 0$, $\beta \neq 0$;  dotted curves:  $\beta_\text{s} \neq 0$, $\beta = 0$.} 
\label{fig:10CC2-nothermal}
\end{figure}

\begin{figure}[hbtp]
\centering
\includegraphics[scale=0.3]{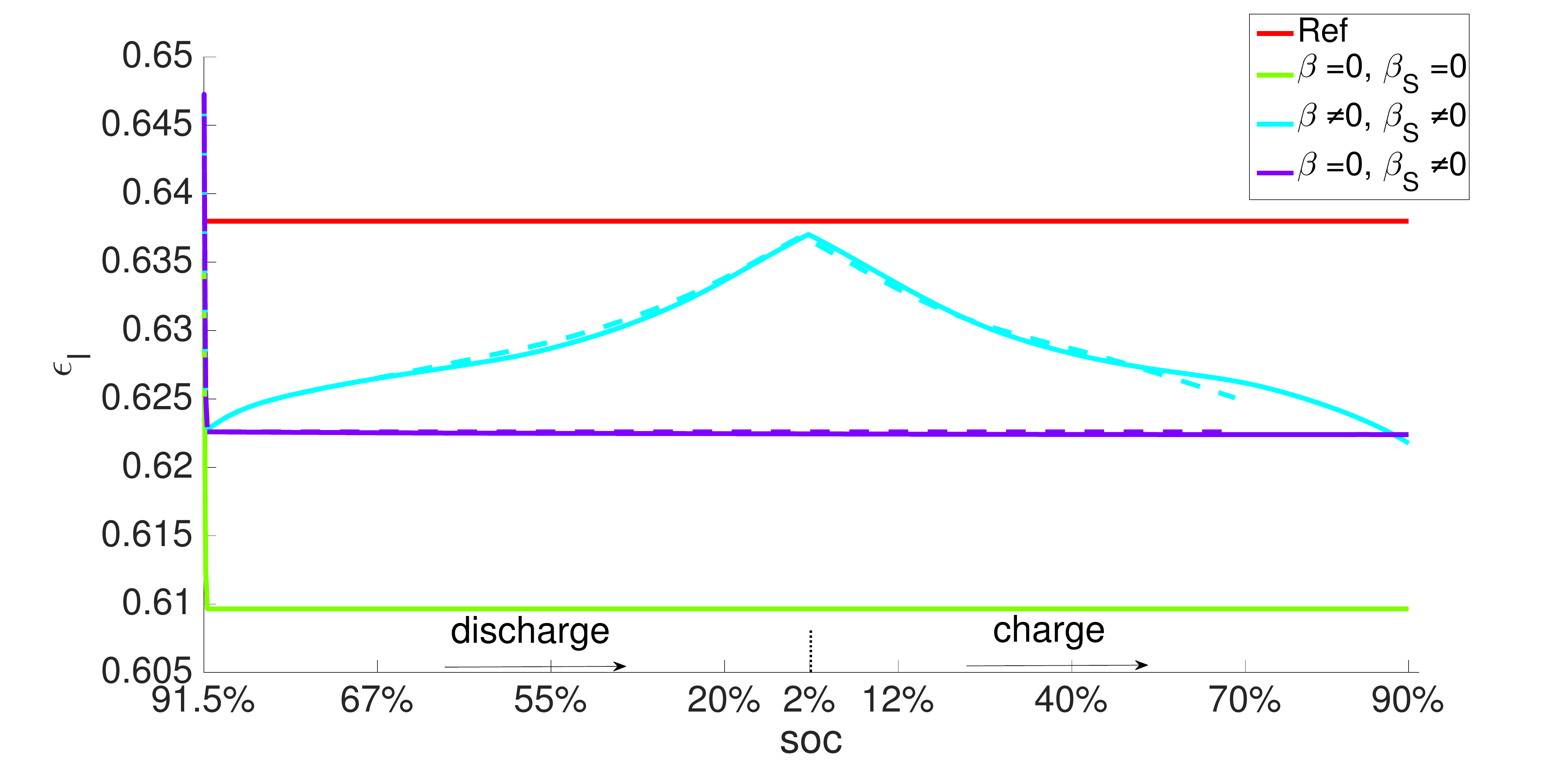}
\caption{{\color{black}Porosity at the midpoint of the separator at the 10~C rate. The midpoint was chosen because the porosity is fairly uniform in the separator (see Figure \ref{fig:10Cepl-thermal}.) }Dashed curves: under isothermal conditions; solid curves: under non-isothermal conditions. The initial drop is due to the displacement boundary condition, which places the cell under compression, as explained in the text. ``Ref'' represents the initial porosity, which would remain fixed were the influence of mechanics not included in the model. From this reference, the porosity falls as soon as the computation starts, due to the compression. } 
\label{fig:10Cepl-sep}
\end{figure}

From the computations, we note that temperature significantly affects the coefficients of the electro-chemical equations; i.e., the transport, conductivity and reaction parameters. Under isothermal conditions, the lithium and lithium ion concentrations are much more non-uniform in the neighborhood of the separator, as seen in Figures \ref{fig:10CC1-nothermal} and \ref{fig:10CC2-nothermal}, and the porosity is also highly non-uniform by the end of the cycle, as seen in Figure \ref{fig:10Cepl-nothermal}. With evolving porosity, the changes in solid particle volume fraction, $\epsilon_\text{s}$,  during discharging/charging exaggerate the nonlinearity of the reaction rate. The reaction is further confined to the region adjacent to the separator at high currents due to the reduced porosity. {\color{black}The non-linear and highly coupled set of equations  (\ref{eq:conserC1Strong}-\ref{eq:BVeuqationU}) which describe the transport and reaction of lithium ions across the cell remains a challenge for numerical solvers as discussed by Ramadesigan et al.\cite{Ramadesigan2012} Our simulations failed to converge after a few time steps at high current rates (10~C), as the local value of $C_1 \rightarrow C_1^\text{max}$ in the negative electrode during charging, and the open circuit potential $U$ approaches zero (see Equations  (\ref{eq:BVeuqation}-\ref{eq:BVeuqationU}) and Figure \ref{fig:U}). In this limit, the surface over-potential $\phi_\text{S}-\phi_\text{E}-U$ takes on large positive values. Consequently, the positive exponential term in Equation (\ref{eq:BVeuqation}) grows, leading to high rates in the Butler-Volmer equation. Combined with the nonlinearly coupled equations, this results in a stiffness due to which the discretized Jacobian matrix is not positive definite even with very small time steps, and the direct solver (UMFPACK) fails. Consequently, we report that at fixed temperature, the computation with constant porosity can be advanced up to SOC = 76\% during charging, which is greater than the maximum SOC of the cases with evolving porosity, as shown in Figure \ref{fig:10CC1-nothermal}. The above trends in the negative electrode are reversed during discharging: $C_1 \rightarrow 0$, $U$ approaches its maximum, $\phi_\text{S}-\phi_\text{E}-U$ takes on large negative values, and the negative exponential term in term in Equation (\ref{eq:BVeuqation}) grows, leading to high rates in the Butler-Volmer equation. In this regime also, the direct solver fails and the numerical solution did not converge after some time.} 

\begin{figure}[hbtp]
\centering
\includegraphics[scale=0.3]{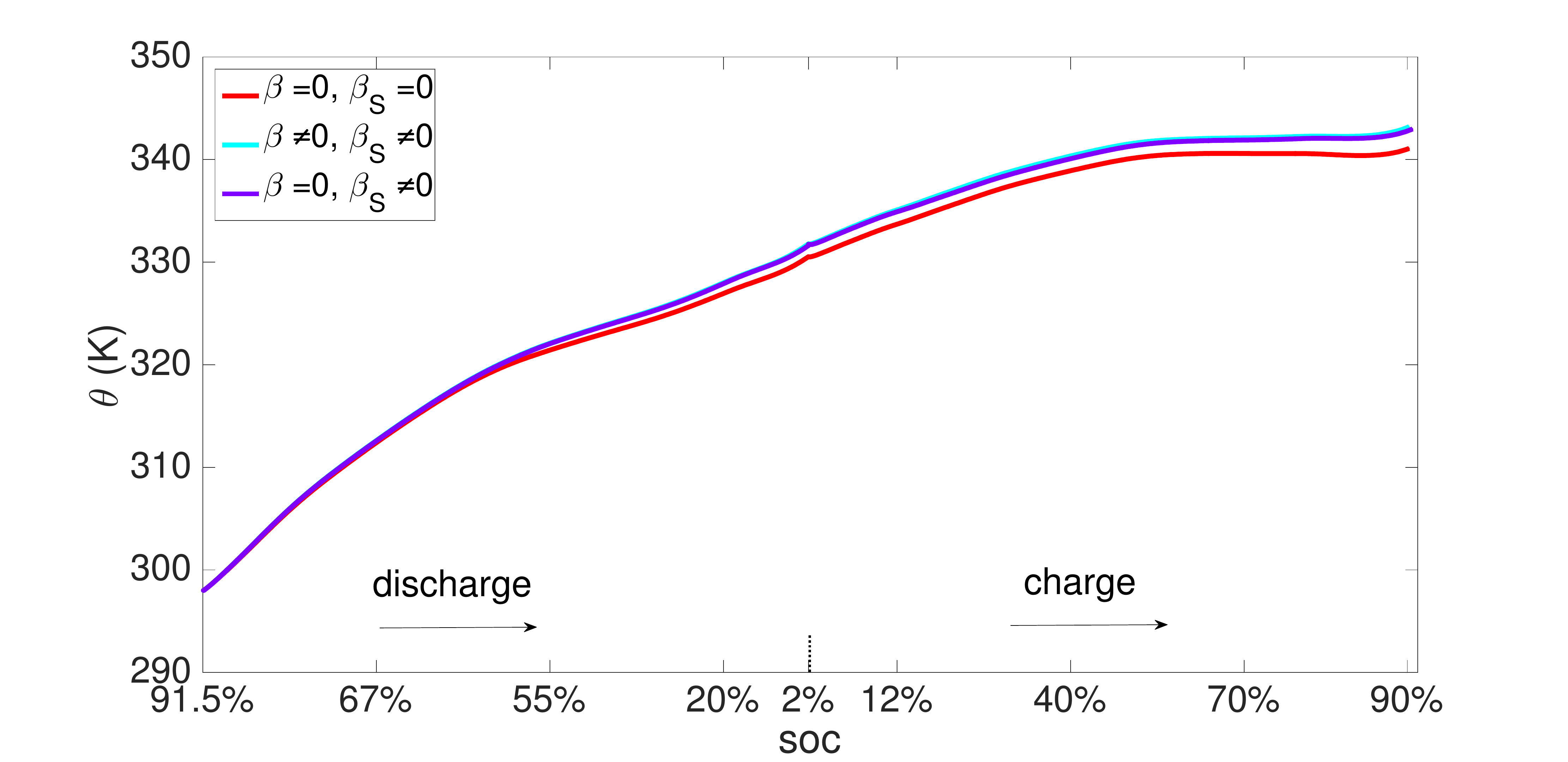}
\caption{The temperature rises by $\sim 45$ K at the 10~C rate during a full discharge-charge cycle.} 
\label{fig:10Ctemp}
\end{figure}

\begin{figure}[hbtp]
\centering
\includegraphics[scale=0.3]{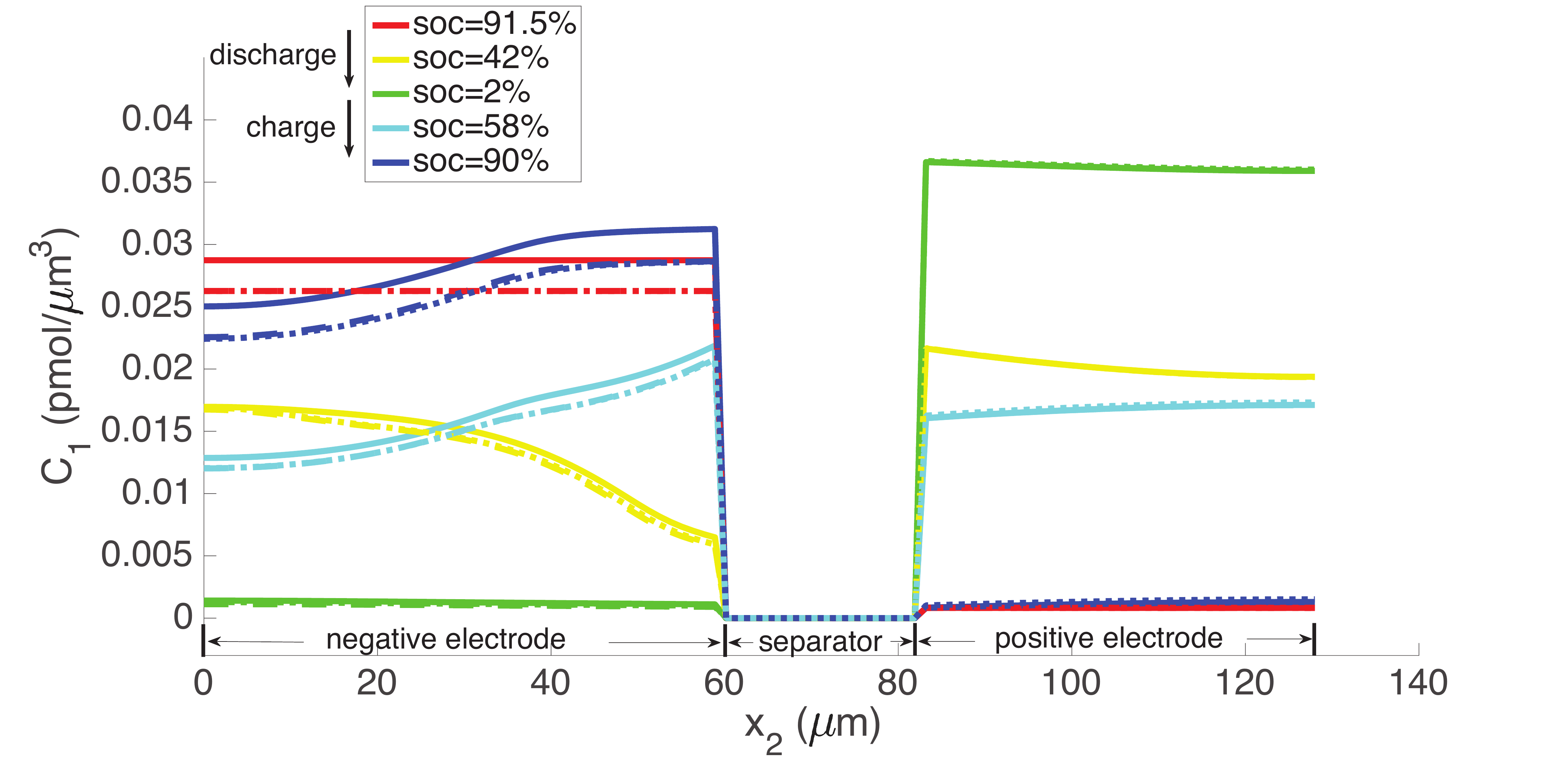}
\caption{Lithium concentration at 10~C rate under non-isothermal conditions. Solid curves: $\beta_\text{s}=0$, $\beta=0$; dashed curves: $\beta_\text{s} \neq 0$, $\beta \neq 0$;  dotted curves: $\beta_\text{s} \neq 0$, $\beta = 0$.} 
\label{fig:10CC1-thermal}
\end{figure}

\begin{figure}[hbtp]
\centering
\includegraphics[scale=0.3]{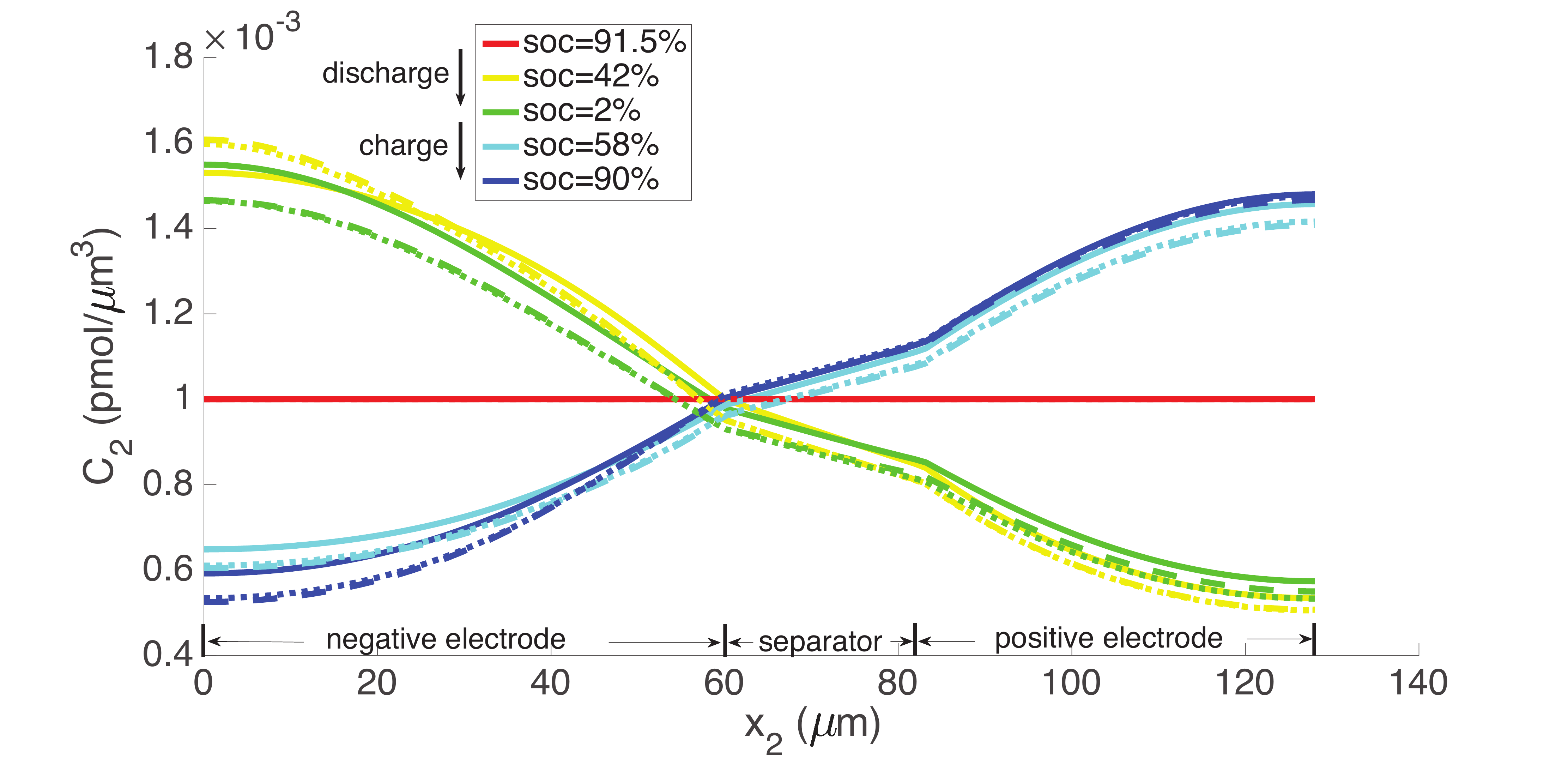}
\caption{Lithium ion concentration at 10~C rate under non-isothermal conditions. Solid curves: $\beta_\text{s}=0$, $\beta=0$; dashed curves: $\beta_\text{s} \neq 0$, $\beta \neq 0$;  dotted curves: $\beta_\text{s} \neq 0$, $\beta = 0$.} 
\label{fig:10CC2-thermal}
\end{figure}

\begin{figure}[hbtp]
\centering
\includegraphics[scale=0.3]{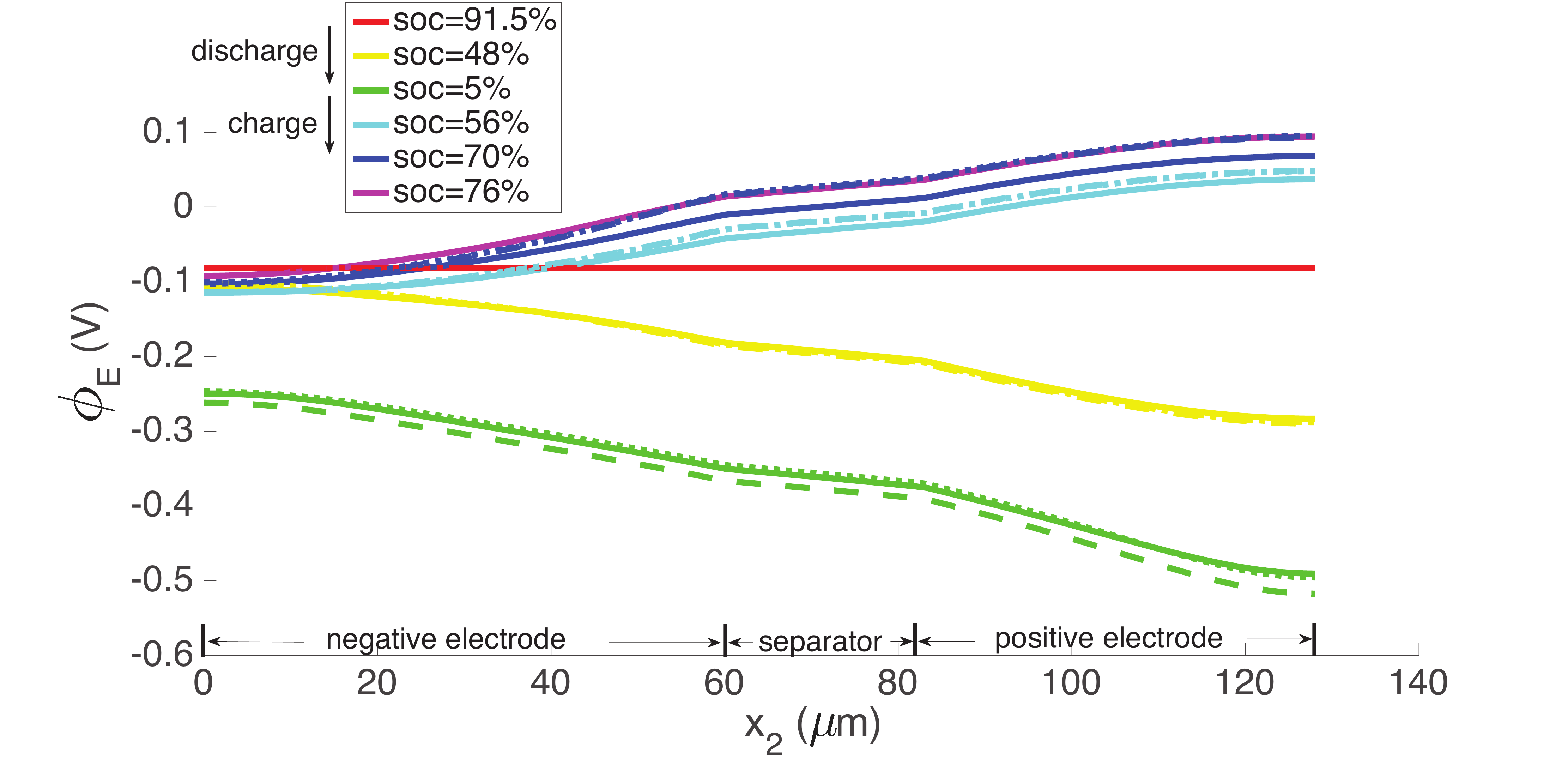}
\caption{The electric potential field in the electrolyte, $\phi_\text{E}$ at 10~C rate under isothermal conditions. Solid curves: $\beta_\text{s}=0$, $\beta=0$; dashed curves: $\beta_\text{s} \neq 0$, $\beta \neq 0$;  dotted curves: $\beta_\text{s} \neq 0$, $\beta = 0$.} 
\label{fig:10Cphie-nothermal}
\end{figure}

\begin{figure}[hbtp]
\centering
\includegraphics[scale=0.3]{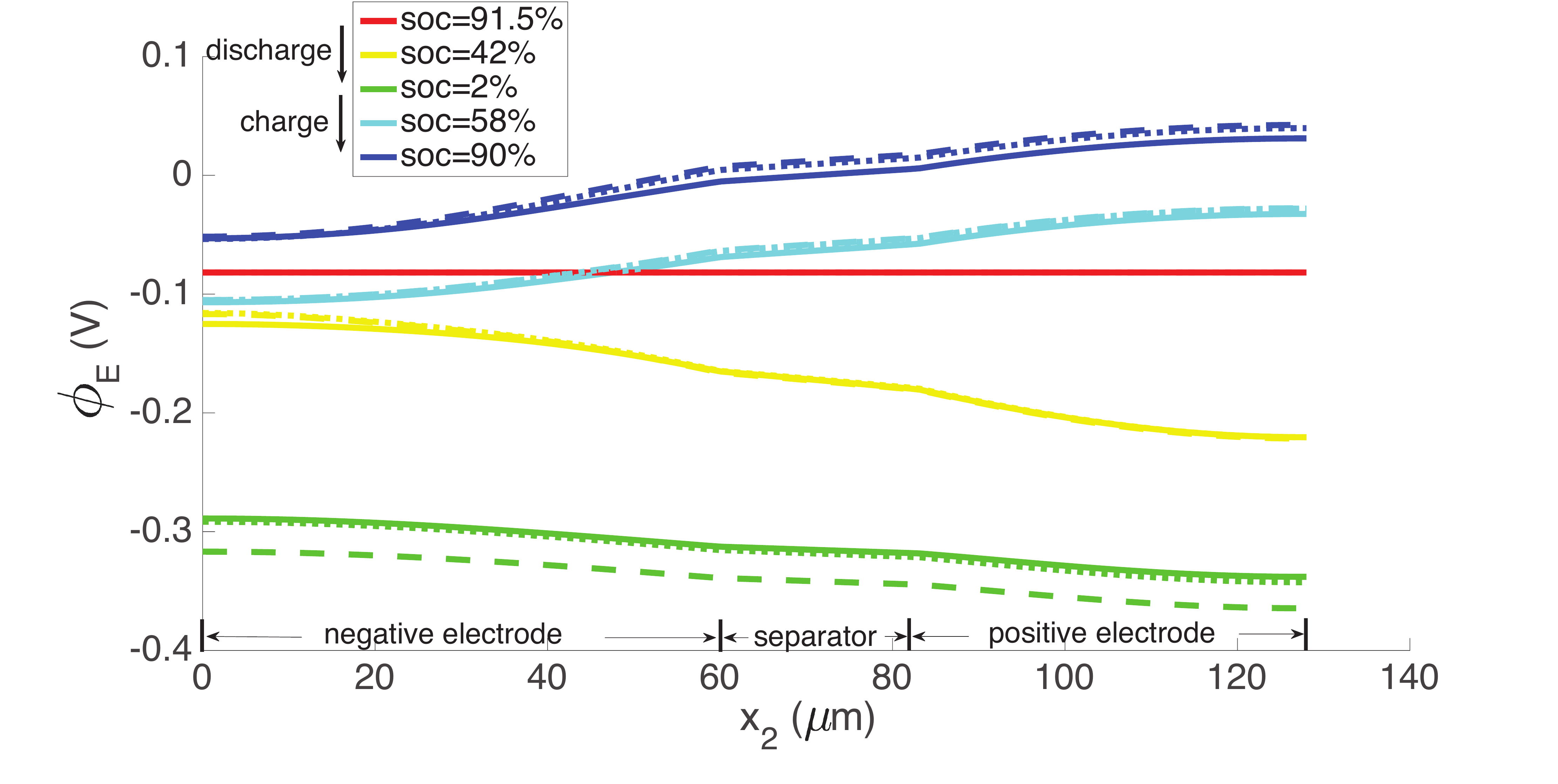}
\caption{The electric potential field in the electrolyte, $\phi_\text{E}$ at 10~C rate under non-isothermal conditions. Solid curve: $\beta_\text{s}=0$, $\beta=0$; dashed curves: $\beta_\text{s} \neq 0$, $\beta \neq 0$;  dotted curves: $\beta_\text{s} \neq 0$, $\beta = 0$.} 
\label{fig:10Cphie-thermal}
\end{figure}

During discharge, the contracting electrode undergoes stress relaxation. Stress equilibrium under the applied displacement boundary condition (Table \ref{tbl:boundaryconditions}) then causes the separator to expand. Thereby, its porosity increases from $\epsilon_\text{l} = 0.62$ to $\sim 0.64$ as seen in the blue curves in Figure \ref{fig:10Cepl-sep}. This porosity gain is surrendered during charging.

{\color{black}As explained above, the high concentration of lithium forces a \textcolor{black}{numerical divergence, which we trace to the growth of exponential terms in the Butler-Volmer model as $C_1 \rightarrow C_1^\text{max}$ during charging, and numerical stiffness of the system of equations causing the direct solver to fail. }\textcolor{black}{ This divergence happens before the cell is fully charged to its initial state from which discharging begins.}} However, this holds only for computations in the isothermal case. During operation, the battery heats up to $\sim 45^\circ$ C at the 10~C rate (Figure \ref{fig:10Ctemp}), leading to higher diffusivities and a more uniform distribution of  lithium concentration as seen by comparing Figure \ref{fig:10CC1-nothermal} with  Figure \ref{fig:10CC1-thermal}, as well as more uniform lithium ion concentrations, as seen by comparing Figure \ref{fig:10CC2-nothermal} and Figure \ref{fig:10CC2-thermal}. The Butler-Volmer model does not \textcolor{black}{cause a numerical divergence of the solution before being fully charged back to its initial state} under these conditions. The porosity also is largely recovered, although at a slightly more non-uniform distribution than before discharge (Figure \ref{fig:10Cepl-thermal}). From the computations we note that the evolving porosity most strongly affects the potential in the electrolyte at states that are close to fully discharged, as seen in Figures \ref{fig:10Cphie-nothermal} and \ref{fig:10Cphie-thermal}. Figure \ref{fig:10Cphis} shows the solid phase potential profile where porosity changes show little effect. However, when the porosity undergoes large changes, such as if the function $\beta_\text{s}(C_1/C^\text{max}_1) \sim 2$ as has been assumed for the non-isothermal case computed in Figure \ref{fig:10Cphis-high-eps}, it has a strong influence on the potential profile. Although such large values of $\beta_\text{s}$ are not the focus of this communication, this scenario may be instructive for studying materials such as tin oxide which undergoes $250\%$ volume expansion\cite{Ebner2013}. The non-symmetric potential profile with respect to discharging/charging in Figure \ref{fig:10Cphis-high-eps} reflects the fundamental irreversibilities in battery operation, here arising due to the transport-reaction and heat conduction/generation phenomena.

During operation, the electrodes deform due to lithium intercalation, thermal expansion and external traction. Since the stress induced by cell deformation is uniform through the cell thickness in these effectively one-dimensional phenomena, \textcolor{black}{we have shown the surface traction force evolving with time in Figure \ref{fig:10Cforce}.} The displacement boundary condition imposes an initial compressive stress on the cell. This compressive stress relaxes since the electrode contracts during discharging, and increases again as the electrode swells due to intercalation during charging. Figure \ref{fig:10Cforce} also shows that the force induced by lithium intercalation is large, but that generated by thermal expansion is small and evolves steadily.

\begin{figure}[hbtp]
\centering
\includegraphics[scale=0.3]{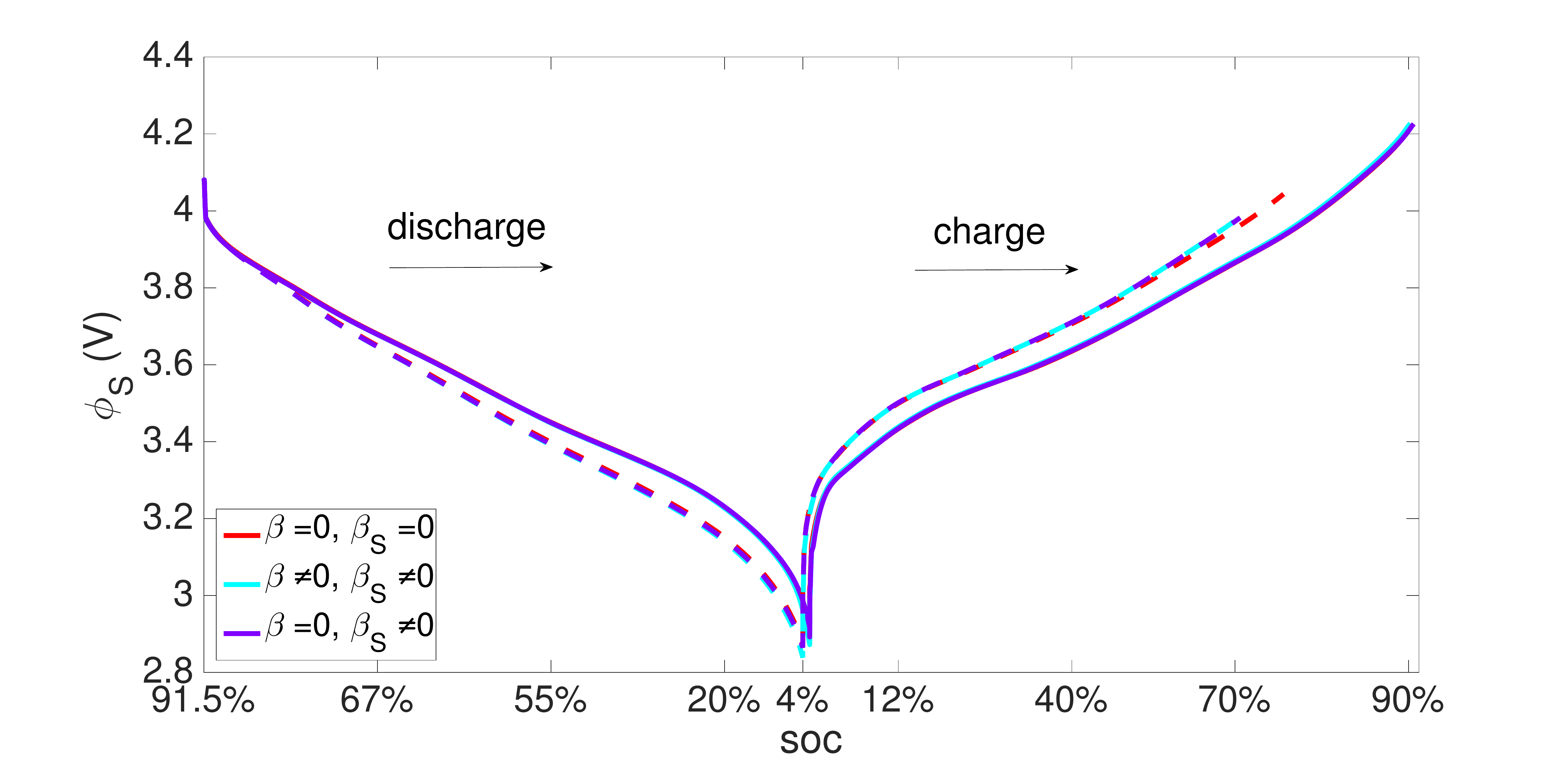}
\caption{$\phi_\text{S}$ at 10~C rate. Dashed curves: under  isothermal conditions; solid curves: under non-isothermal conditions.} 
\label{fig:10Cphis}
\end{figure}

\begin{figure}[hbtp]
\centering
\includegraphics[scale=0.3]{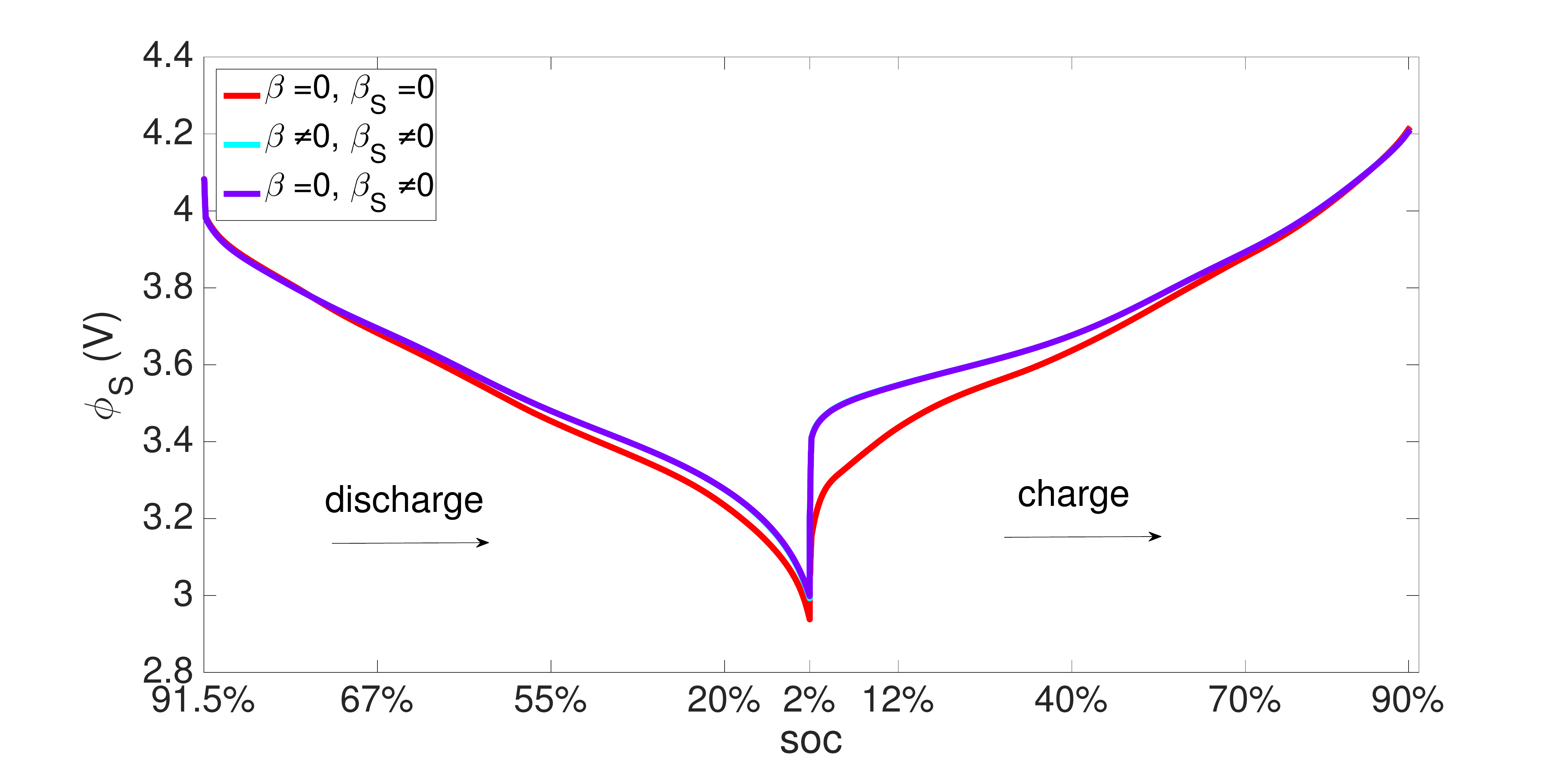}
\caption{$\phi_S$ at 10C rate under non-isothermal conditions with $\sim 200\%$ porosity change: $\beta_\text{s} (C_1/C^\text{max}_1) \sim 2$.} 
\label{fig:10Cphis-high-eps}
\end{figure}

 \begin{figure}[hbtp]
\centering
\includegraphics[scale=0.3]{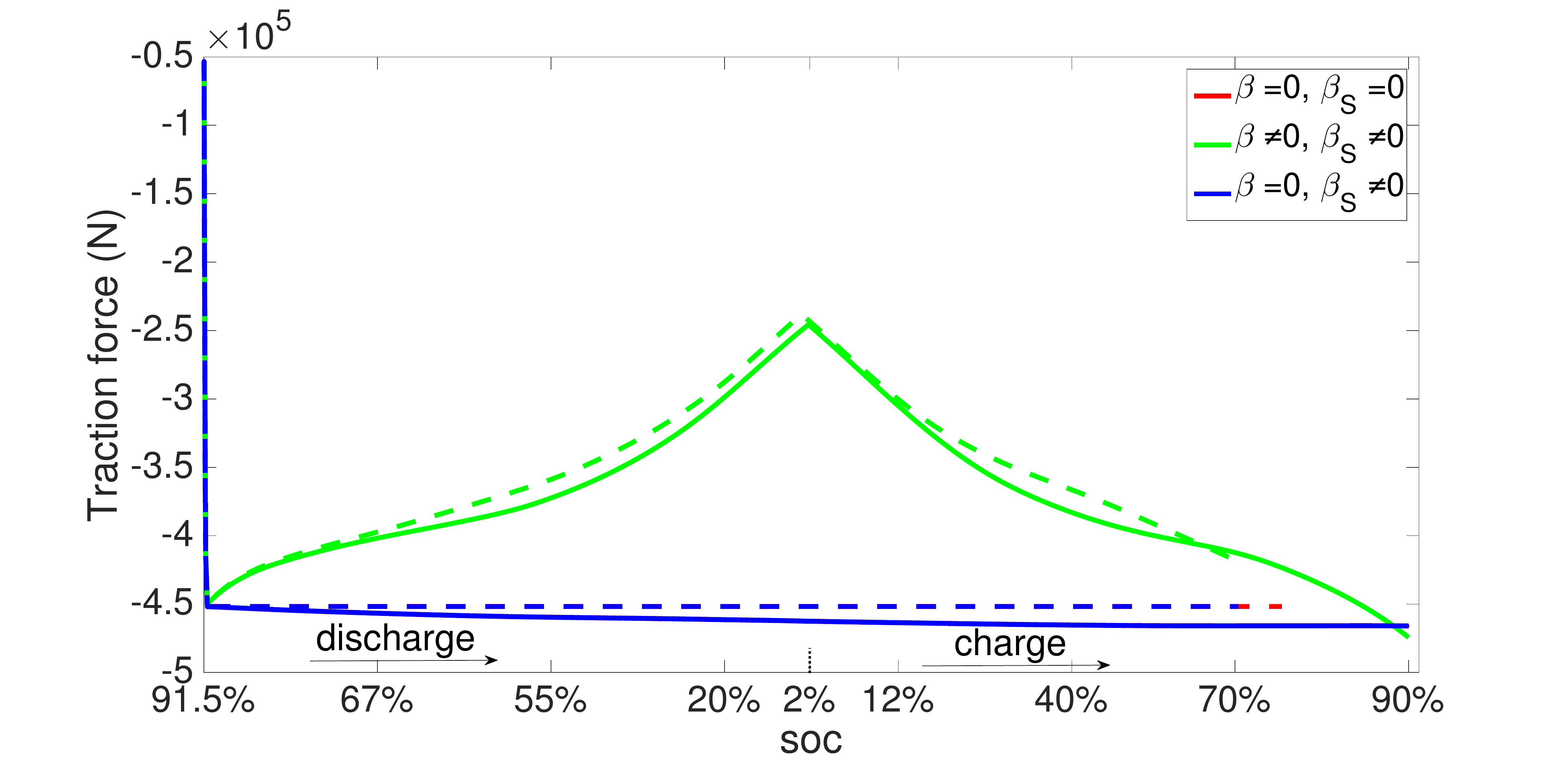}
\caption{Force generated at the 10~C rate. Dashed curves: under  isothermal conditions; solid curves: under non-isothermal conditions. The particle swelling function, $\beta_\text{s}$, is decoupled from the composite electrode swelling function, $\beta$. So, $\beta_\text{s}$ has no influence on the force induced by cell deformation. Accordingly, there is nothing to distinguish between the cases $\beta_\text{s}=0$, $\beta=0$ (red curves) and $\beta_\text{s} \neq 0$, $\beta =0$ (blue curves).}
\label{fig:10Cforce}
\end{figure}

The same initial and boundary value problem was also run at the 1~C current rate under non-isothermal conditions (Figures \ref{fig:1Cepl}-\ref{fig:1Cphis}). Under this low current rate, however, thermal effects are insignificant, the kinetics are slow and thermodynamic dissipation rates are also low. Consequently the battery's performance is much more symmetric in a discharging$\to$charging cycle. Additionally, porosity evolution is uniform, as shown in Figure \ref{fig:1Cepl}, and the lithium concentration is also uniform as shown in Figure \ref{fig:1CC1}. The porosity evolution in the separator in Figure \ref{fig:1C-porosity-sep} and the solid phase potential in Figure \ref{fig:1Cphis} reflect this symmetry. 
\begin{figure}[hbtp]
\centering
\includegraphics[scale=0.3]{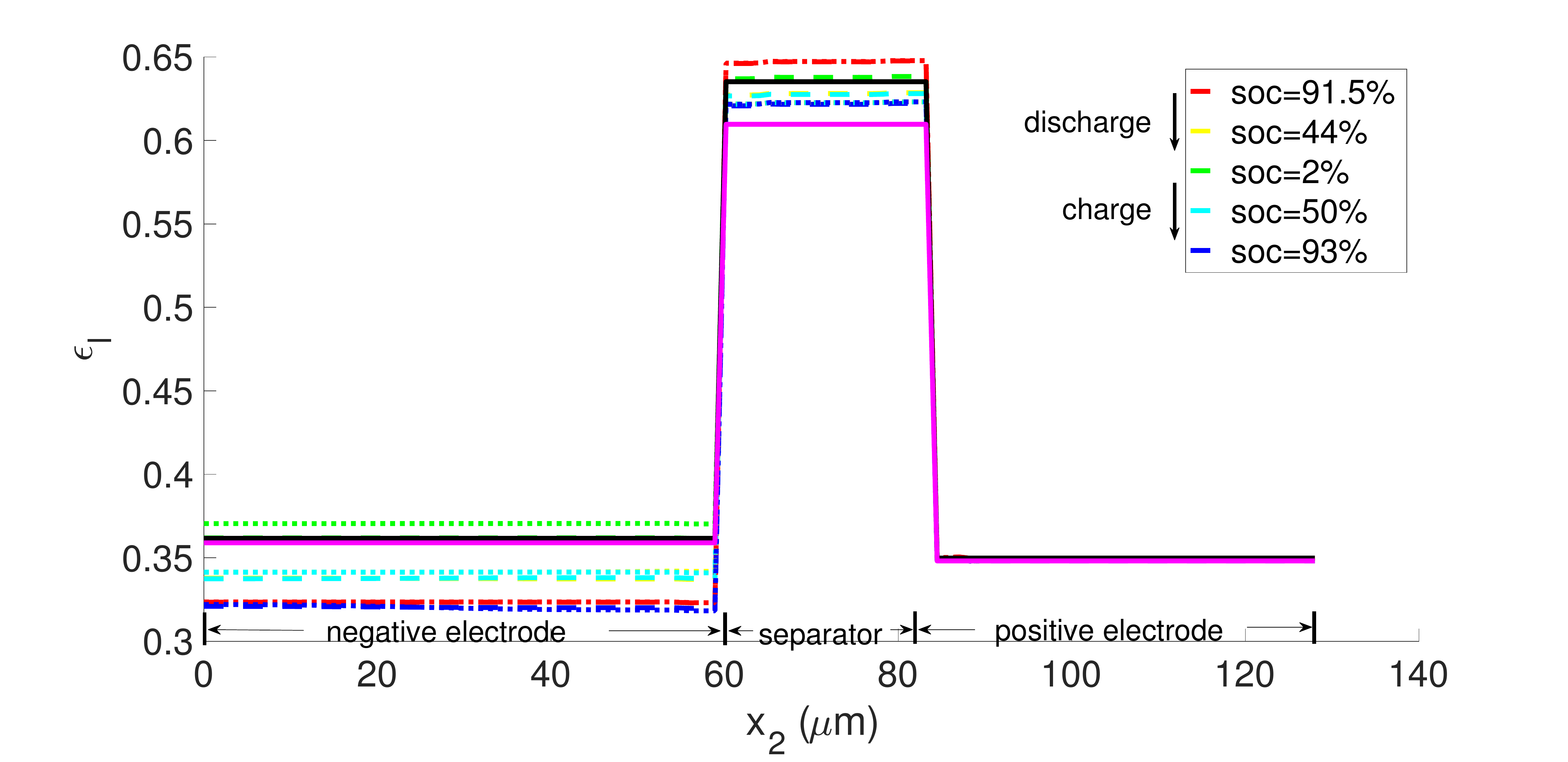}
\includegraphics[scale=0.3]{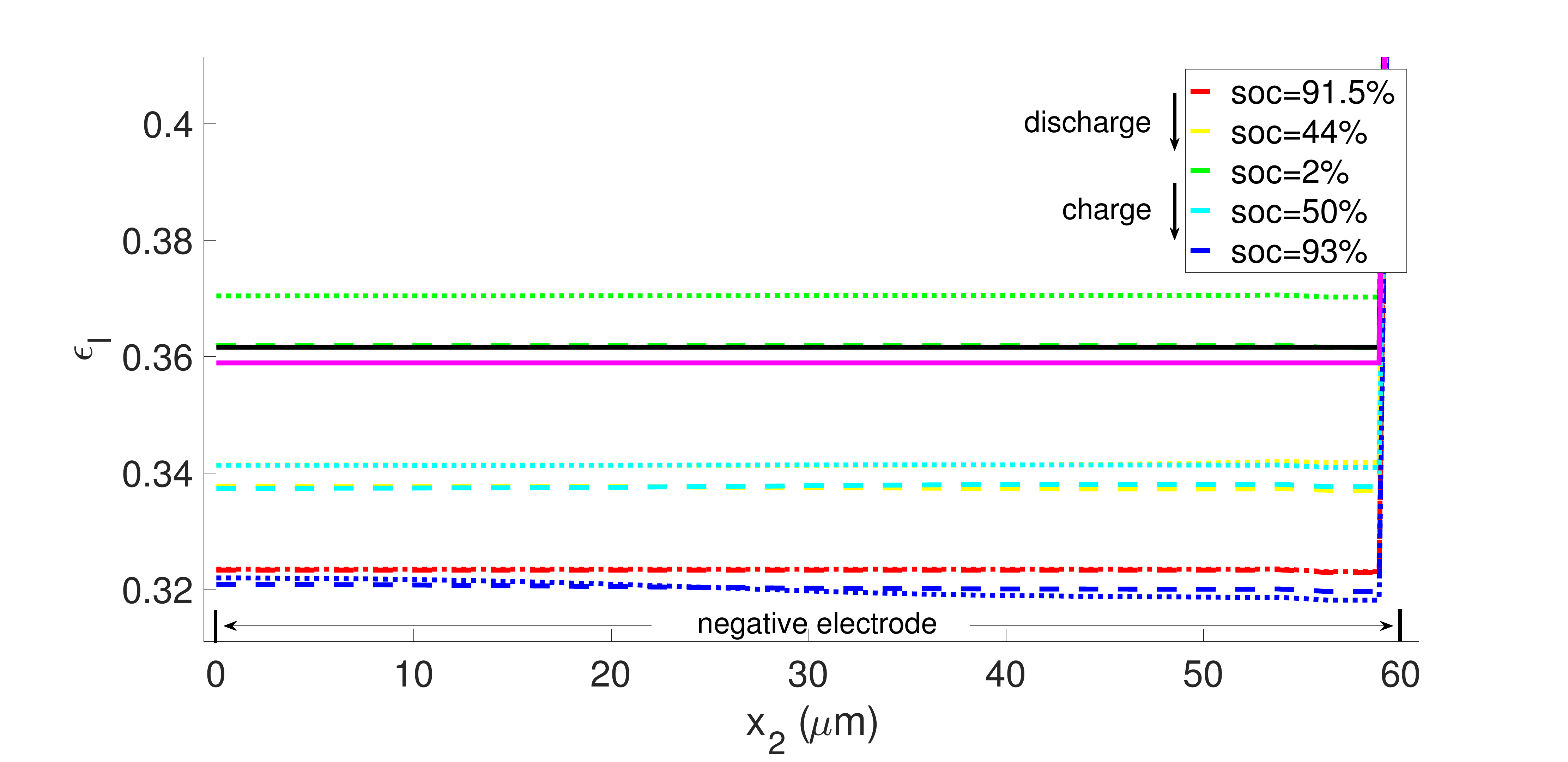}
\caption{Porosity evolving at the 1~C rate. Dashed curves: $\beta_\text{s} \neq 0$, $\beta \neq 0$;  dotted curves: $\beta_\text{s} \neq 0$, $\beta = 0$. Porosity at the low current rate is reasonably uniform. The solid curves represent porosity at the initial state (black) and just after compression (purple).} 
\label{fig:1Cepl}
\end{figure}

\begin{figure}[hbtp]
\centering
\includegraphics[scale=0.3]{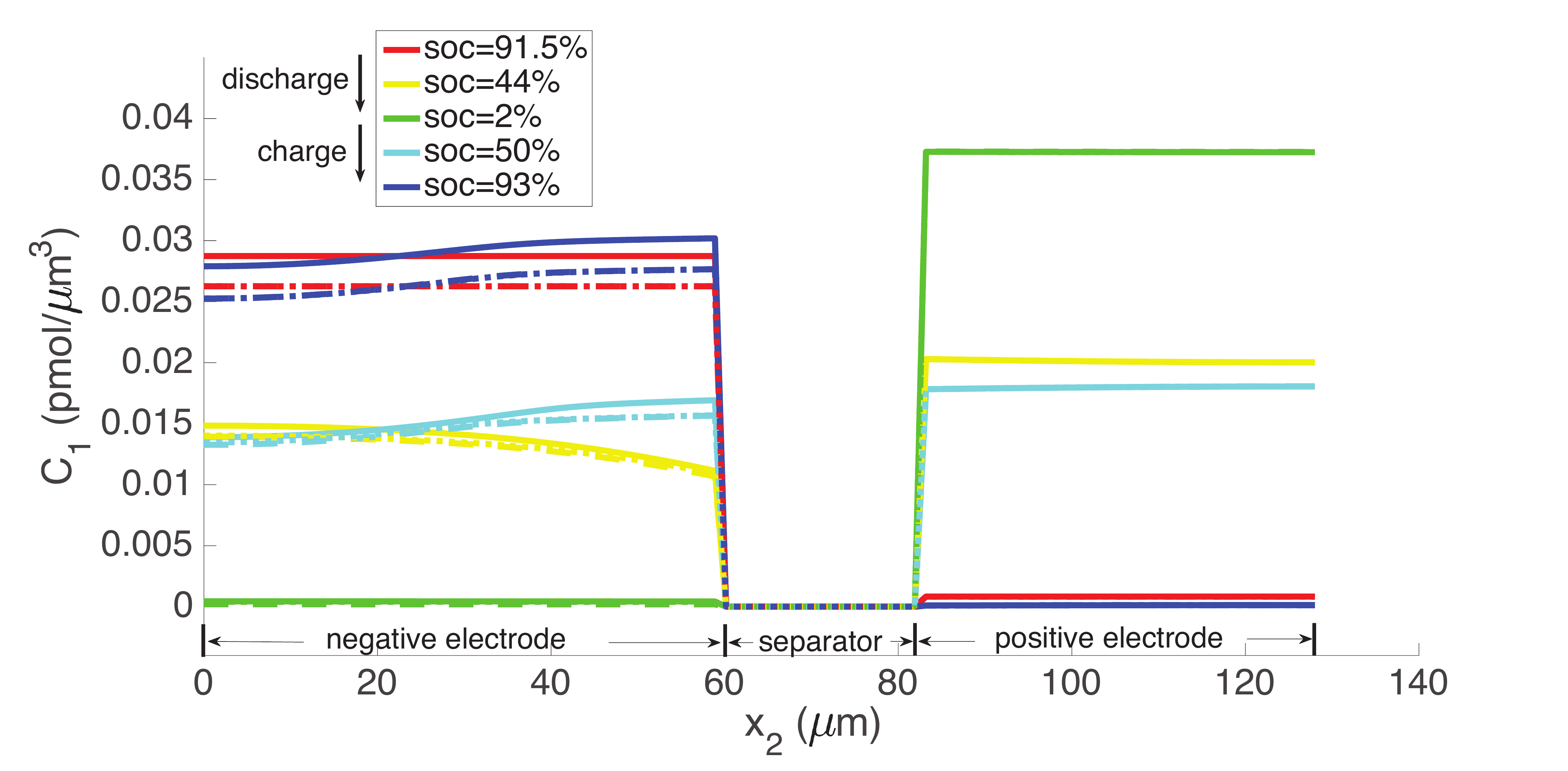}
\caption{Lithium concentration at the 1~C rate. Solid curves: $\beta_\text{s}=0$, $\beta=0$; dashed curves: $\beta_\text{s} \neq 0$, $\beta \neq 0$;  dotted curves: $\beta_\text{s} \neq 0$, $\beta = 0$.} 
\label{fig:1CC1}
\end{figure}

\begin{figure}[hbtp]
\centering
\includegraphics[scale=0.3]{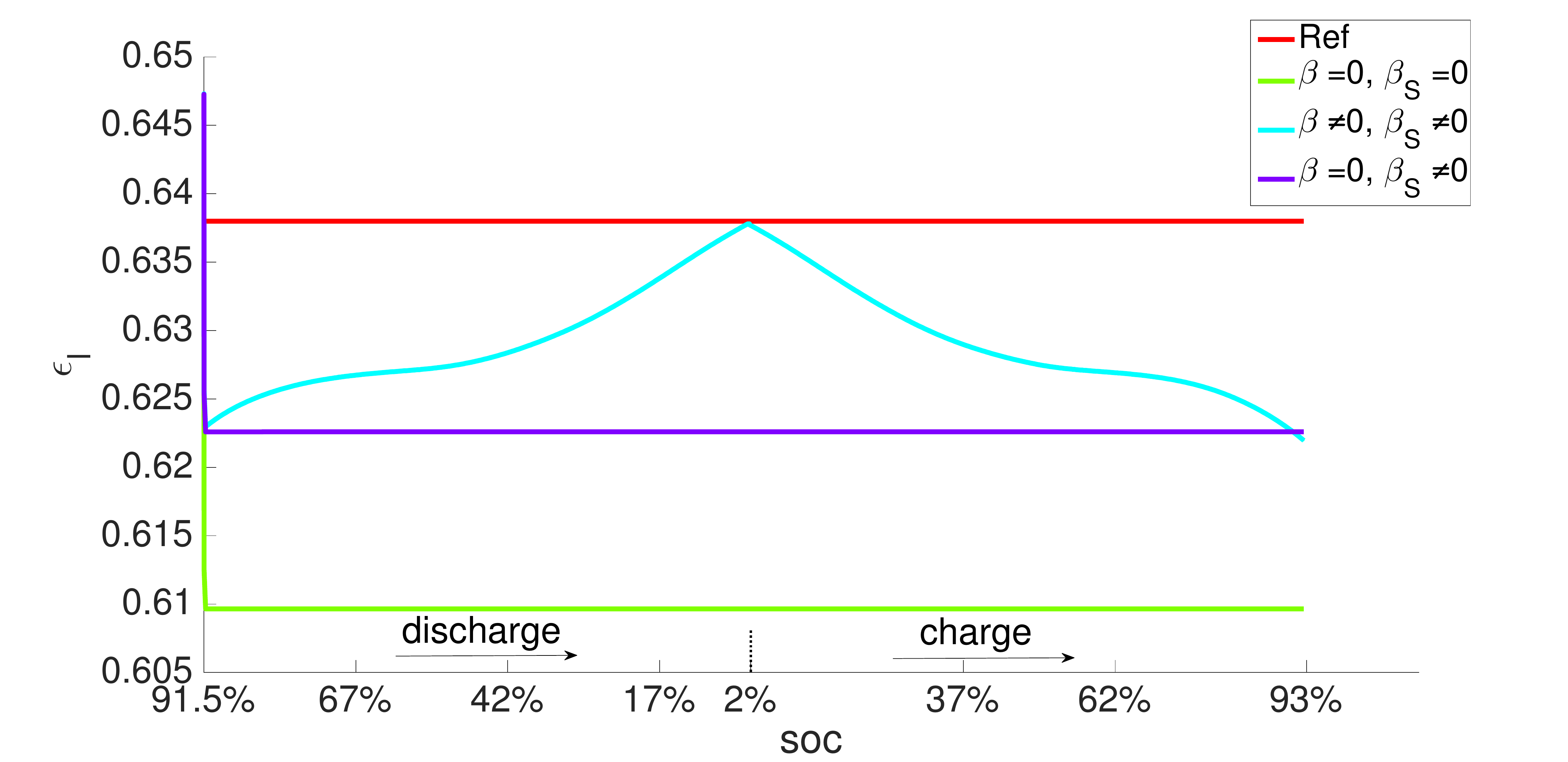}
\caption{Porosity in the separator at the 1~C rate. ``Ref'' represents the initial porosity, which would remain fixed were the influence of mechanics not included. The porosity changes in the separator due to its compression/expansion as the adjacent electrodes deform. As a result, for $\beta=0$, there is no discernible porosity change in  the separator, given that thermal expansion is also insignificant. At this low current rate the small thermal expansion causes the porosity to decrease slightly.} 
\label{fig:1C-porosity-sep}
\end{figure}

\begin{figure}[hbtp]
\centering
\includegraphics[scale=0.3]{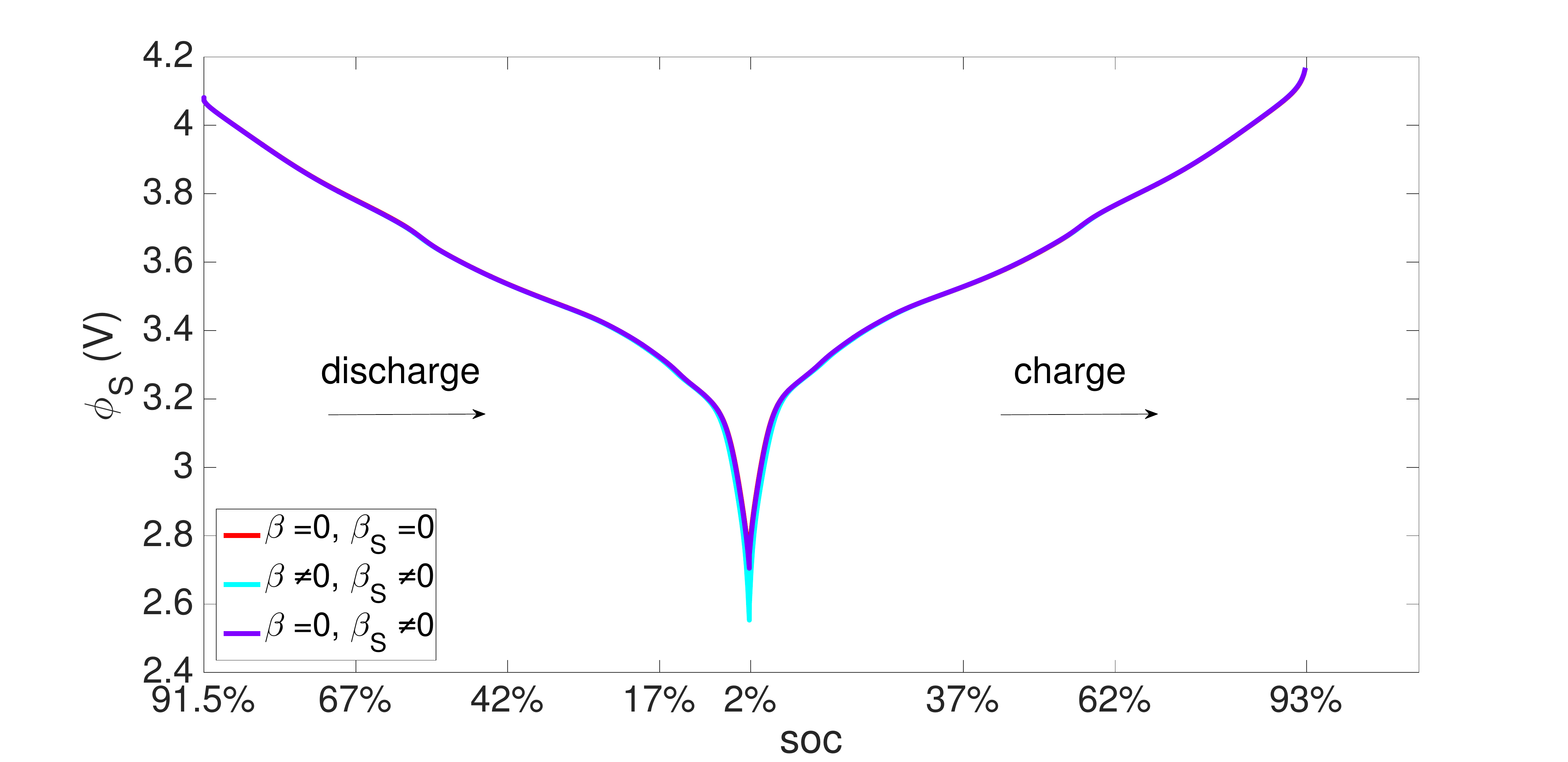}
\caption{The terminal voltage at the 1~C rate, which is given by the electric field in the solid, $\phi_\text{S}$, evaluated at boundary of the domain. At this low C-rate the change in porosity has virtually no impact on the cell terminal voltage.} 
\label{fig:1Cphis}
\end{figure}



\section{Discussion and conclusions}
\label{sec:conclusions}
Porosity studies in the literature on battery materials have focused on modeling fixed microstructures of porous electrodes as outlined in the Introduction. {\color{black}In principle, microscale models can account for the porous microstructure and its evolution. However it has remained hard to bridge scales and to have the microscale evolution be reflected in macroscale simulations of electrodes. 

Non-uniform lithium intercalation and de-intercalation cause active particle expansion and contraction, respectively, thus making the porosity vary as a field over the cell. This variation induces spatially non-uniform stress fields governed by the  partial differential equations of mechanics. The spatial non-uniformity notwithstanding,  porosity induced by lithium intercalation/de-intercalation  is often described in terms of the cell-level quantity of SOC \cite{Awarke2011}. To state the obvious, such approaches ignore spatial variations and, therefore, the results of this non-uniformity. In contrast, our work has taken a step toward a framework in which the dynamics of lithium intercalation/de-intercalation drive the evolution of spatially non-uniform porosity expressed in terms of spatially non-uniform volume fraction fields. Our model has been developed for three dimensional, finite strain elasticity, motivated by the large expansions caused by lithium intercalation (see Figures \ref{fig:beta} and \ref{fig:beta_s}). While based on mixture theory, the model does use a number of simplifications as seen in Section \ref{sec:porosity}. These include the constancy in time and uniformity in space of the pore pressure and stress in the binder, as well as the isotropy of the stress in the binder. However, the stiffness of the particles relative to the composite electrode implies that $\kappa_\text{s} \gg \kappa \det\bF$ in Equation (\ref{voumefractionSolid}), \textcolor{black}{and we assume that $P_\text{l}, P_\text{b} \ll \kappa_\text{s}$.} 
It is therefore apparent from Equation (\ref{voumefractionSolid}) that the solid volume fraction and porosity are controlled by the particle's intercalation-induced swelling and thermal expansion. The preceding assumptions thus do not have a large effect on the computed solutions, but it is desirable to improve upon these models. 

We also draw attention to the fact that, in this first communication, rather than immediately demonstrate the model on three-dimensional domains and the attendant variations in porosity, we have chosen to study our coupled formulation on simpler problems. We have therefore considered effectively one-dimensional initial and boundary value problems for the changes wrought in Li$^+$ ion concentration, Li intercalation, volume fractions, and electric potential, by employing the coupled electro-chemo-thermo-mechanical models. Our main focus in this regard has been on the effects of the evolving porosity.}

Porosity changes affect the coefficients of transport-reaction and electrostatics, and thereby affect battery performance. Lithium intercalation causes swelling and decreases porosity in the electrodes. Swelling electrodes decrease the porosity in the separator {\color{black} by inducing a compressive stress in the cell,} {\color{black} when the total length of the cell is constrained, which is typical of hard-cased prismatic automotive battery cells}. Because of the higher speeds of elastic wave propagation relative to diffusion and migration, the mechanics is solved under quasi-static conditions. External loads therefore affect porosity uniformly and instantaneously, unlike lithium intercalation, which is subject to the rates dictated by kinetics. Thus driven by the mechanics, for the parameter set used in our computations, the porosity decreases more sharply in the separator than in the electrodes due to the low stiffness of the former. For the same parameter set, thermal expansion has little effect on porosity since it is small compared to the swelling caused by lithium intercalation and external loads. However, the temperature strongly affects almost all transport-reaction parameters. 

In our computations the thickness change ($\sim 9\%$) of the electrode caused by lithium intercalation does not have a very strong effect on porosity, but this may not be the case when studying materials such as tin oxide which undergoes $250\%$ volume expansion\cite{Ebner2013}. The accompanying extreme mechanical deformation will have very pronounced effects on porosity, which our model can capture  by proper parametrization of the swelling functions $\beta_\text{s}(C_1)$ and $\beta(C_1)$. {\color{black} \textcolor{black}{Here, we have parametrized the cell swelling functions at low discharge rates where the concentration distribution across the electrode remains uniform, thus enabling correlation with the measurements on the bulk electrode.} This assumption was verified by the 1~C rate simulations which showed uniform concentration distribution across the electrode, whereas the expansion data was collected at the C/20 rate. The model highlights the importance of coupled mechanical deformation, porosity decrease and local transport especially at high rates when the reaction distribution is non-uniform. These local transport limitations from decreased porosity will be especially important for the modeling and design of both high power cell, and high energy density cells with thick electrodes.}

{\color{black} In this computational study, our material data have come from a few different sources. For other parameter sets, we do expect some quantitative changes in the predictions. The main assumption in this regard is the absence of porosity changes in the positive electrode, motivated by the studies of Wang et al. \cite{Wang2012} and Oh et al.\cite{KiYong2014}. For a model that reflects porosity changes of the positive electrode, we would expect qualitatively different results where transport limitations in the cathode would also contribute to the overall terminal voltage response at high current rates. If both electrodes expanded during lithium intercalation, the stress and porosity change in the separator would be reduced, as compared to the case when only one electrode expands, because during discharge or charge one electrode would be expanding while the other contracts.}

{\color{black}We have only fitted $\beta(C_1)$ and $\beta_\text{s}(C_1)$ as functions of SOC, However, $\beta(C_1)$, the expansion/contraction due to lithium intercalation/de-intercalation, depends on the electrode microstructure. Computations that fully resolve the  microstructure at the particle scale would operate with concentration fields varying within the particle, from which properly non-uniform fields would emerge for $\bF^c_\text{s}$, provided that such a model can be parameterized completely. Computational homogenization schemes could then be applied to extract $\beta_\text{s}(C_1)$, which can be defined as an average over the active particles within a representative volume, as well as $\beta(C_1)$. 

The computational homogenization scheme would also apply to other electrode-level parameters such as the diffusivities, conductivities and the pre-factor of the Butler-Volmer equation. The expansion/contraction due to lithium intercalation/de-intercalation in the crystal structure within a particle could be obtained by density functional theory (DFT) computations. Kinetic Monte Carlo methods, with energy barriers obtained by DFT, could provide kinetic parameters for the particle scale computations. In this work we have adopted the Bruggeman relations with constant coefficients to describe the effects of porosity upon transport parameters at the electrode scale. However, these relations also could be replaced by more accurate forms obtained by computational homogenization of particle scale models. Similar paths toward model development appear in the literature \cite{Salvadori2014,Salvadori2015}, and we propose that models such as ours can serve as platforms on which to test the predictions, at the macroscopic scale, of such studies.}

\section*{Acknowledgement}
The information, data, or work presented herein was funded in part by the Advanced Research Projects AgencyEnergy (ARPA-E), U.S. Department of Energy, under Award  \#DE-AR0000269. The computations have been carried out on the Flux computing cluster at University of Michigan, using hardware resources supported by the U.S. Department of Energy, Office of Basic Energy Sciences, Division of Materials Sciences and Engineering under Award \#DE-SC0008637 that funds the PRedictive Integrated Structural Materials Science (PRISMS) Center at University of Michigan.

\subsection*{Disclaimer}
The information, data, or work presented herein was funded in part by an agency of the United States Government. Neither the United States Government nor any agency thereof, nor any of their employees, makes any warranty, express or implied, or assumes any legal liability or responsibility for the accuracy, completeness, or usefulness of any information, apparatus, product, or process disclosed, or represents that its use would not infringe privately owned rights. Reference herein to any specific commercial product, process, or service by trade name, trademark, manufacturer, or otherwise does not necessarily constitute or imply its endorsement, recommendation, or favoring by the United States Government or any agency thereof. The views and opinions of authors expressed herein do not necessarily state or reflect those of the United States Government or any agency thereof.


\appendix
\section{Appendix: Derivation of the electro-chemical equations with finite strain of the electrodes and separator}
The material balance for concentration of lithium $C_1$ in a particle in the deformed configuration is
\begin{align}
\frac{\mathrm{d}}{\mathrm{d} t}\int_{\Omega_\text{p}}C_{1}\mathrm{d}v+\int_{\Gamma_\text{p}}j_n \mathrm{d}s=0
\end{align}
where $j_n$ is the outflux of lithium over the particle, which is integrated over $\Gamma_\text{p}$, surface of the particle. Suppose that there are $N$ particles, of volume $V_\mathrm{p}^i = m(\Omega_\mathrm{p}^i)$, ($i = 1,\dots N$) in a representative element of volume $V_\mathrm{e} = m(\Omega_\mathrm{e})$. Then,
\begin{align}
\sum\limits_{i=1}^N V_\mathrm{p}^i=V_\mathrm{e}\epsilon_\text{s}
\end{align}
where $\epsilon_\text{s}$ is the volume fraction of solid particles. In the representative volume element
\begin{align}
\frac{\mathrm {d}}{\mathrm{d} t}\int_{\Omega_\mathrm{e}}\epsilon_\text{s}C_{1}\mathrm{d}v+\sum\limits_{i=1}^{N}{S_\mathrm{p}^i}j_n^i=0,
\end{align}
where $S_\mathrm{p}^i$ is the surface area of particle $i$, and $j_n^i$ is the normal flux, assumed constant over the particle. It can also be written as 
\begin{align}
\frac{\mathrm{d}}{\mathrm{d} t}\int_{\Omega_\mathrm{e}}\epsilon_\text{s}C_{1}\mathrm{d}v+ \sum\limits_{i=1}^{N}a_\mathrm{p}^i V_\mathrm{p}^i j_n^i=0
\end{align}
where the specific area is $a_\mathrm{p}^i = S_\mathrm{p}^i/V_\mathrm{p}^i$, and reduces to the inverse radius $a_\mathrm{p}^i=3/R_\mathrm{p}^i$ for spherical particles. Note that it varies with change in volume of the solid particle during lithium intercalation or de-intercalation. Replacing the sum with an integral in the limit of a large number of particles in the volume $\Omega_\mathrm{e}$, leads to
\begin{align}
\frac{\mathrm{d}}{\mathrm{d} t}\int_{\Omega_\mathrm{e}}\epsilon_\text{s}C_{1}\mathrm{d}v+\int_{\Omega_\mathrm{e}}\epsilon_\text{s}{a_\mathrm{p}}j_n \mathrm{d}v=0
\end{align}
where quantities in the integrand are regarded as continuous functions of position, $\bx$.
Pulling back the volume integration to the reference configuration by a change of variables,
\begin{align}
\frac{\mathrm{d}}{\mathrm{d} t}\int_{\Omega_{e0}}\epsilon_\text{s}C_{1}J\mathrm{d}V+\int_{\Omega_{e0}}\epsilon_\text{s}{a_\text{p}}j_nJ\mathrm{d}V=0
\end{align}
where $J = \mathrm{det}\bF$ and $\bF$ is the deformation gradient tensor over the porous electrode. Recognizing that $C_1(\bx,t)$ is parameterized in terms of the deformed configuration, and using a standard result for $\dot{J} = J \mathrm{div}\bv$, we have
\begin{align}
\int_{\Omega_{e0}}\left((\frac{\partial (\epsilon_\text{s}C_{1})}{\partial t}+\nabla(\epsilon_\text{s}C_{1})\cdot \bv)J+\epsilon_\text{s}C_{1}\frac{d J}{d t} \right) \mathrm{d}V+\int_{\Omega_{e0}}\epsilon_\text{s}{a_\text{p}}j_nJ\mathrm{d}V=0
\end{align}
For smalll velocities $\bv$ and volume deformation rates $\mathrm{div}\bv$, this reduces to
\begin{align}
\int_{\Omega_{e0}}\frac{\partial (\epsilon_\text{s}C_{1})}{\partial t}J  \mathrm{d}V+\int_{\Omega_{e0}}\epsilon_\text{s}{a_\text{p}}j_nJ\mathrm{d}V=0,
\end{align}
and in the deformed configuration,
\begin{align}
\int_{\Omega_\text{e}}\frac{\partial (\epsilon_\text{s}C_{1})}{\partial t}\mathrm{d}v+\int_{\Omega_\text{e}}\epsilon_\text{s}{a_\text{p}}j_n\mathrm{d}v=0,
\end{align}
The ordinary differential equation for mass balance of lithium is obtained by a standard localization argument:
\begin{align}
\frac{\partial}{\partial t}({\epsilon_\text{s}C_{1}})+\epsilon_\text{s}{a_\text{p}}j_n=0,\quad \mathrm{in} \;\Omega_{e}
\label{eq:App-conserC1Strong}
\end{align}

The material balance for lithium cations in the polymer/salt electrolyte $C_2$ is,
\begin{align}
\frac{\partial C_2}{\partial t}=-\nabla\cdot \bN_+-\frac{s_+}{nF}\nabla\cdot \bi_2
\label{eq:App-c2}
\end{align}
where $\bN_+$, the  flux of cations homogenized over both matrix and pore, is
\begin{align}
\bN_+=C_2\bv^*- \epsilon_\text{l}D_\text{eff}\nabla C_2+\frac{t^0_+}{z_+nF}\bi_2
\end{align}
with $\bv^*$ being the velocity averaged over matrix and pore. The  current density $\bi_2$ in the pore phase is related to pore wall flux $j_n$ in the electrolyte phase through the relation
\begin{align}
a_\text{p}Fj_n=\nabla \cdot \bi_2
\label{eq:App-reactioneq}
\end{align}
Equations (\ref{eq:App-c2})  to (\ref{eq:App-reactioneq}) are in the form used by Newman and Tiedemann\cite{Newman1975}. 
Consistent with the assumption of low material velocities, we set $\bv^* = \mathbf{0}$ and $t^0_+$ to be a constant. Written over the electrode volume in the deformed configuration, the integral form of lithium cation balance is:
\begin{align}
\frac{\mathrm{d}}{\mathrm{d} t}\int_{\Omega_{e}} \epsilon_\text{l}C_{2}\mathrm{d}v=\int_{\Omega_e}-\nabla\cdot(-\epsilon_\text{l}D_\text{eff}\nabla C_{2})\mathrm{d}v-\int_{\Omega_{e}}\left(\frac{t^0_+}{z_+n}+\frac{s_+}{z_+n}\right)\epsilon_\text{s}{a_\text{p}}j_n \mathrm{d}v
\label{eq:App-intc2}
\end{align}
Commonly for lithium-ion batteries the choice $\frac{t^0_+}{z_+n}+\frac{s_+}{z_+n}=-1+t^0_+$ is made\cite{Moyle1993}, leading to:

\begin{align}
\frac{\mathrm{d}}{\mathrm{d} t}\int_{\Omega_{e}} \epsilon_\text{l}C_{2}\mathrm{d}v=\int_{\Omega_e}\nabla \cdot(\epsilon_\text{l}D_\text{eff}\nabla C_{2})\mathrm{d}v+(1-t^0_+)\int_{\Omega_e}\epsilon_\text{s}{a_\text{p}}j_n \mathrm{d}v
\end{align}
Rewriting the above equation by pulling back to the reference configuration, we have
\begin{align}
\frac{\mathrm{d}}{\mathrm{d} t}\int_{\Omega_{e_0}} \epsilon_\text{l}C_{2}J \mathrm{d}V=\int_{\Omega_{e_0}}\nabla \cdot(\epsilon_\text{l}D_\text{eff}\nabla C_{2})J \mathrm{d}V+(1-t^0_+)\int_{\Omega_{e_0}}\epsilon_\text{s}{a_\text{p}}j_nJ \mathrm{d}V
\end{align}
Persisting with the assumption of low velocities $\bv$ and volumetric rates of deformation  $\mathrm{div}\bv$, this leads to
\begin{align}
\int_{\Omega_{e_0}}\frac{\partial (\epsilon_\text{l}C_{2})}{\partial t}J  \mathrm{d}V=\int_{\Omega_{e_0}}\nabla \cdot(\epsilon_\text{l}D_\text{eff}\nabla C_{2})J \mathrm{d}V+(1-t^0_+)\int_{\Omega_{e_0}}\epsilon_\text{s}{a_\text{p}}j_nJ \mathrm{d}V
\end{align}
and, over the deformed electrode configuration,
\begin{align}
\int_{\Omega_\text{e}}\frac{\partial (\epsilon_\text{l}C_{2})}{\partial t}J  \mathrm{d}v=\int_{\Omega_\text{e}}\nabla \cdot(\epsilon_\text{l}D_\text{eff}\nabla C_{2}) \mathrm{d}v+(1-t^0_+)\int_{\Omega_\text{e}}\epsilon_\text{s}{a_\text{p}}j_n \mathrm{d}v
\end{align}
A standard localization argument leads to the partial differential equation for mass balance of $\text{Li}^+$ ions:
\begin{align}
\frac{\partial }{\partial t}(\epsilon_\text{l}C_{2})=\nabla\cdot (\epsilon_\text{l}D_\text{eff}\nabla C_{2})+(1-t^0_+)\epsilon_\text{s}{a_\text{p}}j_n
\label{eq:App-conserC2Strong}
\end{align}
Here we have begun the derivation of the equations in the deformed configuration and pulled them back to the reference configuration, mainly because the actual transport coefficients are typically measured over samples that have not undergone large deformation under the effect of lithiation and thermal expansion. The final equations, however, are in the deformed configuration.

We next consider the equations for the electric fields in the form presented by Doyle et al.\cite{Moyle1993}. The current density in the solid phase, $\bi_1$, is governed by Ohm's law:
\begin{align}
\bi_1=-\sigma_\mathrm{eff}\nabla\phi_\text{S}
\label{App-eq:currenti1}
\end{align}
where $\sigma_\mathrm{eff}$ is the effective conductivity and $\phi_\text{S}$ is the potential in the solid phase.
The current density in the electrolyte phase, $\bi_2$, is driven by the gradient of the corresponding potential, $\phi_\mathrm{E}$ as well as the cation potential gradient:
\begin{align}
\bi_2=-\gamma_\mathrm{eff}\nabla \phi_\text{E}+\frac{\gamma_\mathrm{eff}R\theta}{F}(1+\frac{\partial\ln f}{\partial \ln C_2})(\frac{t^0_+}{z_+n}+\frac{s_+}{z_+n})\nabla\ln C_2
\label{App-eq:currenti2}
\end{align}
where $\theta$ is temperature and $k_\mathrm{eff}$ is the effective conductivity corresponding to $\phi_\text{E}$.
Gauss' Law takes on the form
\begin{align}
\nabla\cdot\bi_1 &= -a F j_n\\
\nabla\cdot\bi_2 &= a F j_n
\label{App-eq:gausslaw}
\end{align}
where $F$ is the Faraday constant. Note that the above equations guarantees current conservation over the effective electrode:
\begin{equation}
\nabla \cdot(\bi_1+\bi_2)=0
\label{App-eq:currentConserve}
\end{equation}
Re-arranging equations (\ref{eq:App-reactioneq}) and (\ref{App-eq:currenti1}--\ref{App-eq:currentConserve}), and using $ 1+\frac{\partial\ln f}{\partial \ln C} \approx 2$ \cite{Pollard1980}, we obtain
\begin{equation}
\nabla\cdot\left(\gamma_\text{eff}(-\nabla\phi_\text{E})+\gamma_\text{eff}\frac{2R\theta}{F}(1-t^{0}_{+})\uline{\nabla}\ln C_{2}\right)=aFj_{n}
\label{App-eq:phieEq}
\end{equation}
\begin{equation}
\nabla\cdot\left(\sigma_\text{eff}(-\nabla\phi_\text{S})\right)=-aFj_{n}
\label{App-eq:phisEq}
\end{equation}

Finally, we detail the derivation between the intra-particle lithium concentration, $c_1$ and its value averaged over the particle, $C_1$. Starting with the assumption of a parabolic profile for $c_1$,
\begin{align}
c_1=a_0+a_1r+a_2r^2
\end{align}
The profile satisfies an  influx boundary condition at the particle's surface 
\begin{align}
D_\text{s}\frac{\partial c_1}{\partial r}\bigg\vert_{r=R}=-j_n\label{eq:CsBCsurface}
\end{align}
where $D_\text{s}$ is the diffusivity of lithium within the particle. Symmetry at $r  = 0$ requires
\begin{equation}
\frac{\partial c_1}{\partial r}\bigg\vert_{r=0}=0. 
\label{eq:Cssymm}
\end{equation}
The local concentration, volume-averaged over the particles, is the field denoted by $C_1$. It is obtained by solving Equation (\ref{eq:conserC1Strong}), and by using
\begin{align}
C_1=\frac{1}{V_\text{p}}\int_0^{R_\text{p}}c_1 4\pi r^2dr
\label{eq:aveC1withCs}
\end{align}
Solving Equation (\ref{eq:aveC1withCs}) with Equations (\ref{eq:Cssymm}) and (\ref{eq:CsBCsurface}) we can determine the three coefficients in the parabolic profile as
\begin{align}
a_0=C_1-\frac{3a_2R_\text{p}^2}{5};\quad    a_1=0; \quad a_2=\frac{-j_n}{2D_\text{s}R_\text{p}}
\end{align}
Then the surface concentration is obtained as Equation (\ref{eq:C1surface}).

\section{Appendix: Material data}
{\color{black} The swelling and voltage data describing a graphite negative electrode, microporous polypropylene/polyethylene separator and lithium nickel/manganese/cobalt oxide (NMC) positive electrode were used for this study. A full set of electrochemical model parameters for NMC was not available for this study, therefore baseline parameters from the literature (from other cathode materials where appropriate) were used as the initial point for simulation. Identification and validation of the parameters against experimental data for the Carbon-NMC cell will be investigated in a following communication. Since the positive electrode expansion is minimal the graphite properties were critical for visualizing the impact on cell performance due to electrode swelling as highlighted in Fig~\ref{fig:10Cepl-thermal}.

{\color{black}Upon defining $\eta=\frac{C_1}{C^{\text{max}}_1}$, the function  $\frac{\partial U}{\partial \theta}$ can be written as a fit with the following forms, and is shown in Figure \ref{fig:dUdT}:
\begin{equation}
\frac{\partial U}{\partial \theta}=\left.
\begin{cases}
0.01442*\eta^2-0.00291*\eta-0.000138 & \eta<=0.2 \\
0.00634*\eta^3-0.006625*\eta^2+0.002635*\eta-0.0004554 & 0.2<\eta<=0.4  \\
0.001059*\eta-0.0004793  & 0.4<\eta<=0.5  \\
0.00025*\eta-7.5\times10^{-5}  & 0.5<\eta<=0.7  \\
-0.001*\eta+0.0008  & 0.7<\eta<=0.8  \\
0.0333*\eta^2-0.057*\eta+0.02427 & 0.8<\eta<=0.82  \\
0.002*\eta^2-0.0039*\eta+0.00177 & 0.82<\eta<=0.95  \\
-0.0014*\eta+0.0012& 0.95<\eta<=1
\end{cases}
\right.
\label{eq:dUdT}
\end{equation}

\begin{figure}[hbtp]
\centering
\includegraphics[scale=0.3]{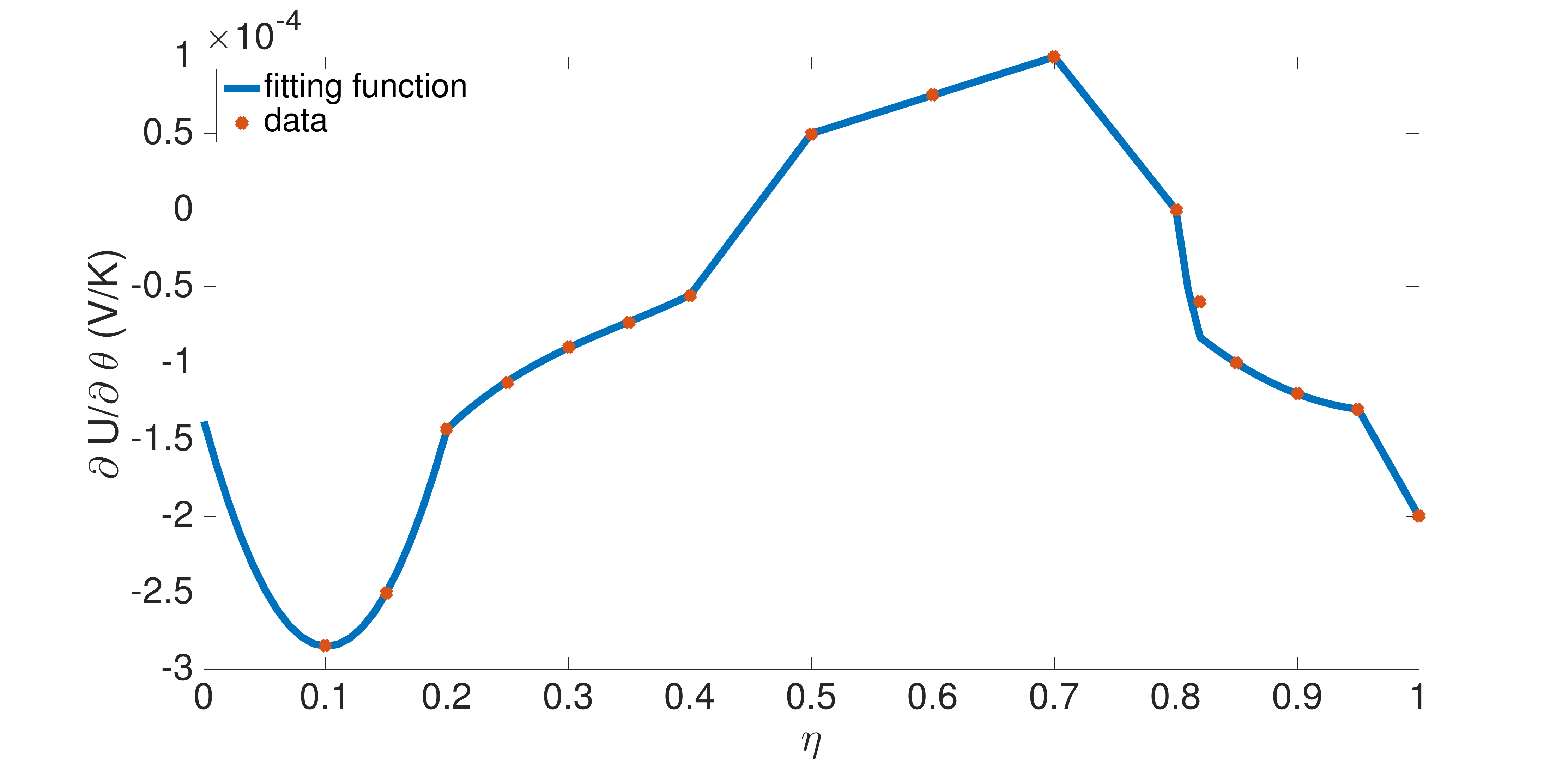}
\caption{Data for $\partial U/\partial \theta$ is from Oh et al.\cite{KiYong2014}. The solid curve is a fit given by Equation (\ref{eq:dUdT}). Here, $ \eta=\frac{C_1}{C_1^\text{max}}$.}
\label{fig:dUdT}
\end{figure}
}

\begin{longtable}{c|c|c|c|c|c}
 \hline
Symbol &  Name&Unit & LiC$_6$ & Sep & NMC\\ \hline
 & constant & \multicolumn{4}{c}{}\\ \hline
$F$ & Faraday's constants  &  pC/pmol &\multicolumn{3}{c} {96487}\\ \hline
$R$ & Universal gas constant & pJ/(pmol$\cdot$K)& \multicolumn{3}{c}{8.3143}\\ \hline
$\theta_0$& Initial temp &K& \multicolumn{3}{c} {298} \\ \hline
&Cell geometry& \multicolumn{4}{c}{}\\ \hline
$l $ &Thickness\cite{KiYong2014} & $\mu$m &60&23 &{45}\\ \hline
$w_1$ &Side length\cite{KiYong2014}  &$\mu$m&\multicolumn{3}{c}{$120\times 10^3$}\\ \hline
$w_2$ &Side length\cite{KiYong2014}  &$\mu$m&\multicolumn{3}{c}{$85\times 10^3$} \\ \hline
 \\ \hline
 & Electrochem parameters &\multicolumn{4}{c}{}\\ \hline
$\alpha_a$ & Transfer coeff\cite{Fuller1994} &-& 0.5 &-&0.5\\ \hline
$\alpha_c$ & Transfer coeff\cite{Fuller1994} &-& 0.5 &-&0.5\\ \hline
$k_0$ & kinetic rate constant\footnote{\label{App:papers}\color{black}Estimated based on references\cite{White2010JES, LimEA2012, Danner2016}.} &$\sqrt{\text{pmol}}/(\mu \text{m}^2\text{s})$& $8.0\times 10^{-4}$ & -& $8.0\times 10^{-4}$\\ \hline
$R_0$ & Radius of solid particles\footref{App:papers}& $\mu$m& 8.0&-& 6.0  \\ \hline
$\sigma$ & Conductivity: active matl\footref{App:papers}& $\text{p}(\Omega \mu \text{m})^{-1}$& $1.5\times 10^8$& -& $0.5\times 10^8$  \\ \hline
$D_\text{s}$ & Diffusivity of lithium \cite{Fuller1994}  & $\mu \text{m}^2/\text{s}$& $5\times 10^{-1}$&-& $1\times 10^{-1}$\\ \hline
 $t_0^+$ & Transfer number\cite{WangJES2007} &- &-&0.2 &-\\ \hline
 $\epsilon_{\text{s}_0}$ &Init solid vol frac\footnote{\label{App:initial}\color{black}Initial conditions.}&-& 0.53 & 0.35 & 0.5 \\ \hline
$\epsilon_{\text{l}_0}$ &Initial porosity\footref{App:initial} &-& 0.32  & 0.65& 0.35\\ \hline
$\epsilon_{\text{s\_r}_0}$ &Init solid vol frac\footref{App:initial} &-& 0.485 &0.362& 0.5\\ 
 &  for no porosity change & && &\\ \hline
$\epsilon_{\text{l\_r}_0}$ &Initial porosity\footref{App:initial} &-& 0.362& 0.638 & 0.35\\
 &  for no porosity change & && &\\ \hline
$C^{\text{max}_0}_1$ & Maximum Li conc\cite{Kim2011} & $\text{pmol}/\mu \text{m}^3$ & $28.7\times 10^{-3}$ &-& $37.5\times 10^{-3}$ \\ \hline
$C_{1soc\_ 100}$ & Maximum SOC\footref{App:initial}&-& 0.915 &-& 0.022\\ \hline
$C_{1soc\_ 0}$ & Minimum SOC\footref{App:initial}&-& 0.02 &-& 0.98 \\ \hline
$C_{2\_ ini}$ & Init conc Li ion\footref{App:initial} & $\text{pmol/}\mu \text{m}^3$&-& $1.0\times 10^{-3}$ &- \\ \hline
 \\ \hline
 &Therm parameters &  \multicolumn{4}{c}{}\\ \hline
$ \rho$& Density\cite{Northrop2015} & $\text{kg}/\mu \text{m}^3$& $2.5\times 10^{-15}$ & $1.1\times 10^{-15}$ & $2.5\times 10^{-15}$\\ \hline
$C_p$ & Specific heat\cite{Northrop2015} & $\text{pJ}/(\text{kgK})$& $7\times 10^{14}$& $7\times 10^{14}$& $7\times 10^{14}$   \\ \hline
$\lambda$&Therm conductivity\cite{Sun2015} &$\text{W/(mK)}$ & $1.04\times 10^6$  & $0.33\times 10^6$& $5\times 10^6$\\ \hline
$h$ & heat transfer coeff\cite{Xiao2011JPS} & $W/(m^2K)$& 5& 5 & 5\\ \hline
$\Omega$ &Therm exp coeff \cite{Xiao2011JPS}& $1/\text{K}$ & $9.615\times 10^{-6}$ & $82.46\times 10^{-6}$& $6.025\times 10^{-6}$\\ \hline
$\Omega_\text{s}$ &Therm exp coeff\footnote{\color{black}Estimated based on properties of carbon.}  & $1/\text{K}$ & $6\times 10^{-6}$ & $6\times 10^{-6}$& $6\times 10^{-6}$ \\ 
& of AM particle&&&&\\ \hline
 \\ \hline
&Elasticity parameters& \multicolumn{4}{c}{}\\ \hline
$\kappa$& Bulk modulus: cell\footnote{\label{App:bulk}\color{black}Calculated by its Young's modulus.}& GPa & $4.94\times 10^{-3}$ & $0.42\times 10^{-3}$ & $7.4\times 10^{-3}$\\ \hline
$\kappa_\text{s}$ & Bulk modulus: carbon\footref{App:bulk}& GPa & $25\times 10^{-3}$&-& $25\times 10^{-3}$ \\ \hline
$E$& Young's modulus: cell\cite{Xiao2011JPS} &GPa& 5.93 & 0.5& 8.88 \\ \hline
$\nu$ & Poisson's ratio\cite{Roberts2016} &-& 0.3& 0.3& 0.3\\ \hline
$u_0$& Disp boun cond\footnote{\color{black}Fitted to data\cite{KiYong2014}.}&$\mu$m& \multicolumn{3}{c}{-0.24}\\ \hline
\end{longtable}
}


\bibliographystyle{unsrt}

\end{document}